\documentclass[journal=jacsat,manuscript=article]{achemso}

\usepackage[version=3]{mhchem} % Formula subscripts using \ce{}

\usepackage{graphicx}% Include figure files
\usepackage{dcolumn}% Align table columns on decimal point
\usepackage{bm}% bold math
\usepackage[utf8]{inputenc}
\usepackage[T1]{fontenc}
\usepackage{mathptmx}
\usepackage{amsmath,amssymb}
\usepackage{bm}
\usepackage{cleveref}
\usepackage{booktabs}
\usepackage{notes2bib}
\usepackage{braket}
\usepackage{xcolor}
\usepackage[english]{babel}

\usepackage{verbatim}

\makeatletter
\def\@email#1#2{%
 \endgroup
 \patchcmd{\titleblock@produce}
 {\frontmatter@RRAPformat}
 {\frontmatter@RRAPformat{\produce@RRAP{*#1\href{mailto:#2}{#2}}}\frontmatter@RRAPformat}
 {}{}
}%

\DeclareUnicodeCharacter{2212}{-}
\DeclareUnicodeCharacter{2061}{}

\def\br{\mathbf{r}}

\def\h2o{\mathrm{H}_2\mathrm{O}}

\def\icomp{\mathfrak{i}}

\author{Arno Förster}
\email{a.t.l.foerster@vu.nl}
\affiliation{Theoretical Chemistry, Vrije Universiteit Amsterdam, De Boelelaan 1105, 1081 HV Amsterdam, The Netherlands}%

\author{Fabien Bruneval}
\email{fabien.bruneval@cea.fr}
\affiliation{Universit\' e Paris-Saclay, CEA, Service de recherche en Corrosion et Comportement des Matériaux, SRMP, 91191 Gif-sur-Yvette, France}

\title{Why does the $GW$ approximation give accurate quasiparticle energies? The cancellation of vertex corrections quantified}


\begin{document}

\begin{abstract}
Hedin's $GW$ approximation to the electronic self-energy has
been impressively successful to calculate quasiparticle energies,
such as ionization potentials, electron affinities, or electronic band structures.
The success of this fairly simple approximation has been ascribed to
the cancellation of the so-called vertex corrections that go beyond $GW$.
This claim is mostly based on past calculations using vertex corrections within
the crude local-density approximation.
Here, we explore a wide variety of non-local vertex corrections in the
polarizability and the self-energy, using first-order approximations or infinite summations to all orders.
In particular, we use vertices based on statically screened interactions like in the Bethe-Salpeter equation.
We demonstrate on realistic molecular systems that the two vertices
in Hedin's equation essentially compensate.
We further show that consistency between the two vertices is crucial
to obtain realistic electronic properties.
We finally consider increasingly large clusters and extrapolate that our conclusions
would hold for extended systems.
\end{abstract}

Spectroscopic properties of many-electron systems are often described in terms of effective equations for single- and two-particle Green's functions first formulated by Hedin. \cite{Hedin1965} Hedin's equations start from the Dyson equation for the single-particle Green's function $G$ and express the corresponding self-energy $\Sigma$ in terms of a dynamically screened electron-electron interaction $W$ and a vertex function $\Gamma$. 

Practical calculations have to approximate the vertex function. The most drastic of these approximations is the $GW$ approximation (GWA),\cite{Hedin1965, Aryasetiawan1998, Golze2019}
where the vertex function is reduced to delta functions.
First applied to extended systems
\cite{Hedin1965,lundqvist_pkm1967,strinati_prl1980,  Strinati1982a, Hybertsen1985, Hybertsen1986, Godby1988, Schone1998} and later to small metal clusters,\cite{Onida1995, Ishii2001, Ishii2002} and molecules \cite{Rohlfing2000a, Grossman2001, Tiago2006, Rostgaard2010, Blase2011, Ke2011, Bruneval2012, Korzdorfer2012} it is by now widely used to describe quasiparticle (QP) levels and band structures in systems as diverse as complex molecules, \cite{Duchemin2012, Forster2022c, Allen2024} molecule-metal interfaces, \cite{Thygesen2009, Liu2019, Adeniran2021, Zhang2023a} dye-sensitized solar cells, \cite{Umari2013, Marsili2017, Belic2022} or Moiré materials. \cite{Brooks2020, Romanova2022, Graml2024}

In weakly correlated systems, the GWA is relatively accurate for two reasons. First, the dynamical screening of the electron-electron interaction at large distances captures a significant source of electron correlation. \cite{Kaasbjerg2010, VanLoon2021} While this seems natural in extended systems, it is remarkable that the GWA often gives highly accurate QP energies in atoms and molecules with sometimes only a few electrons. \cite{Bruneval2012, Bruneval2013, Vacondio2022, Marie2024} This hints towards major cancellations between higher-order terms in the self-energy as a second reason for the success of the $GW$ approximation. However, despite numerous studies, \cite{Mahan1989, Shirley1993, Bobbert1994, delSole1994, DeGroot1996, Shirley1996, Bechstedt1997, Schindlmayr1998, Albrecht1998, Ummels1998, Schindlmayr1998a, Takada2001, Bruneval2005, Tiago2006, Morris2007, Shishkin2007, Gruneis2014, Stefanucci2014, Ren2015, Chen2015, Kutepov2016, Pavlyukh2016, Hung2016, Kuwahara2016, Kutepov2017a, Schmidt2017, Maggio2017, Maggio2018, Kutepov2017, Cunningham2018, Olsen2019, Vlcek2019, Lewis2019, Pavlyukh2020, Wang2021, Mejuto-Zaera2021, Tal2021, Joost2022a, Mejuto-Zaera2022a, Forster2022, Rohlfing2023, Lorin2023, Vacondio2024, Tal2024, Abdallah2024, Wen2024, Patterson2024, Bruneval2024} these cancellations are still poorly understood. The partial cancellation of vertex corrections in $W$ and in $\Sigma$ has first been demonstrated for aluminum \cite{Mahan1989} and silicon. \cite{Bobbert1994, delSole1994, Bruneval2005} While vertex corrections improve fully self-consistent $GW$ (sc$GW$) band gaps and satellites, \cite{Shirley1996, Pavlyukh2016, Kutepov2016, Kutepov2017, Kutepov2017a, Kutepov2021c, Kutepov2022, Rohlfing2023} almost all $GW$ calculations replace the interacting $G$ by an effective non-interacting $G^{(0)}$
that may be judiciously chosen to achieve high accuracy for molecular QP energies. \cite{Knight2016, Bruneval2021a, McKeon2022, Bruneval2024} Many authors argue that partial cancellations of vertex corrections in $W$ and $\Sigma$ in combination with the QP approximation to $G$ are another reason for the success of the GWA in practice \cite{Kotani2007} but this subject is debated. 

% . and explains why inclusion of the vertex in $\Sigma$ only is not a successful strategy if a qualitatively correct $G^{(0)}$ is used \cite{Bruneval2024}.

Including the very same vertex consistently in $W$ and $\Sigma$ allows one to quantify these cancellations rigorously. Following this strategy, previous work has demonstrated the Hartree-Fock (HF) vertex to improve over $GW$ QP excitations and satellites in atoms and small molecules. \cite{Maggio2017, Mejuto-Zaera2021, Vacondio2024} The resulting self-energy is self-screening free \cite{romaniello_jchemphys2009,Chang2012} but comes with the disadvantage that its beyond-$GW$ contribution is expanded in terms of the bare Coulomb interaction instead of the screened one. Especially in larger systems where screening effects are potentially strong, a screened TDHF vertex should be more realistic. Patterson has recently performed such calculations, \cite{Patterson2024,Patterson2024a} albeit within the Tamm-Dancoff approximation (TDA) in $L$ and $\Sigma$. Within the TDA the same vertex has also been used by Cunningham \textit{et al.} \cite{Cunningham2018, Cunningham2023} within quasi-particle self-consistent $GW$ (qs$GW$) \cite{VanSchilfgaarde2006,Faleev2004, Kotani2007, bruneval_springer2014} but without any vertex correction in $\Sigma$. 

We here build on these works and further explore the maze of vertex corrections,
which is still today mostly unmapped.
Our quantitative conclusions are based on well-established
molecular benchmarks where accurate wavefunction method-based results offer
unambiguous references.
We consistently include bare and screened exchange vertices in $W$ and $\Sigma$. 
The TDA is known to be a severe approximation in RPA-based $GW$ calculations \cite{Bintrim2021} and we avoid it here. 
For a wide range of molecules, including one-dimensional and two-dimensional models of graphene and passivated silicon clusters, we demonstrate far-reaching cancellations of vertex corrections, rationalizing the success of the $GW$ approximation from small molecules to extended systems. 

\begin{figure}
    \centering
    \includegraphics[width=\linewidth]{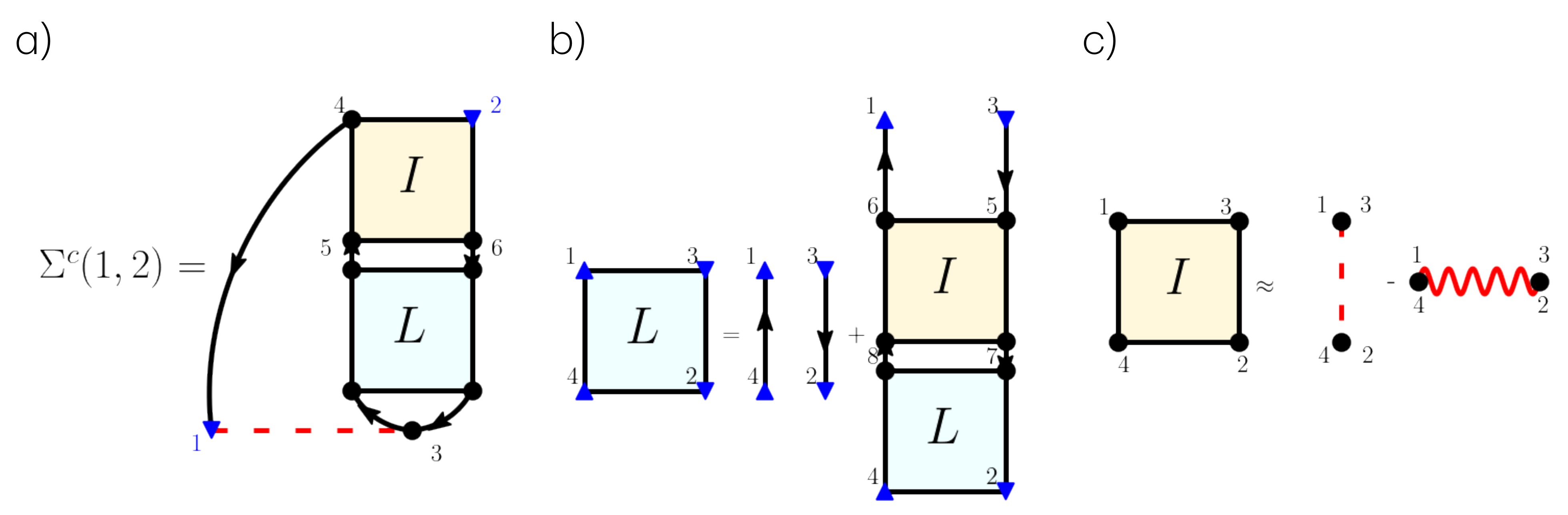}
    \caption{Diagrammatic representation of a) the correlation part of the exact self-energy, b) the two-particle correlation function $L$, and c) the approximate kernel we use in this work. Dotted lines denote the Coulomb interaction $v$, and the wiggly line is the statically screened Coulomb interaction $W_0$.}
    \label{fig:diagrams}
\end{figure}

As shown in Figure~\ref{fig:diagrams}a), we write the self-energy in the form: \cite{Strinati1988, Romaniello2012, Orlando2023}
\begin{equation}
\label{sigma}
\begin{aligned}
\Sigma_{xc}(1,2) = & \icomp v(1^+,2) G(1,2) \\
  & + \icomp v(1^+,3) G(1,4) I(4,6,2,5) L(5,3,6,3) \;,
\end{aligned}
\end{equation}
where integers $n = (\bm{r}_n, \sigma_n, t_n)$ collects spatial coordinates, spin and time, $v$ is the usual 2-point Coulomb interaction and $I(1,2,3,4) = \icomp \delta \Sigma(1,3) / \delta G (4,2)$ is the 4-point irreducible kernel. Integration over repeated indices is implied. In the following, we focus on closed shells only and therefore assume spin compensation. As shown in Figure~\ref{fig:diagrams}b), the two-particle correlation function $L$ is obtained through the solution of a Bethe-Salpeter equation (BSE)
\begin{equation}
\label{BSE}
\begin{aligned}
L(1,2,3,4) = & L^{(0)}(1,2,3,4) \\
    & + L^{(0)}(1,5,3,6) I (6,7,5,8)
    L(8,2,7,4) \;,
\end{aligned}
\end{equation}
where we introduced the non-interacting correlation function $L^{(0)}(1,2,3,4) = -\icomp G(1,4)G(2,3)$ and
the very same kernel $I$ as in $\Sigma$ appears. Complemented with the Dyson equation for $G$, \cref{sigma,BSE} yields a self-consistent scheme that is completely equivalent to Hedin's equations. \cite{Starke2012} The present scheme has the major advantage that the 3-point vertex $\Gamma$ only appears implicitly and its explicit calculation is effectively replaced by the solution of the BSE~in Eq.~\eqref{BSE}. 

We follow previous work \cite{Maggio2017, Vlcek2019, Lewis2019, Bruneval2021a, Mejuto-Zaera2021, Vacondio2024} and exclusively work with a Hartree--Fock (HF) Green's function
\begin{equation}
  \label{eq:g0}
  G^{(0)}(\br,\br',\omega) =
    \sum_i^\mathrm{occ}
     \frac{\varphi_i(\br)   \varphi_i(\br^\prime)}
     {\omega - \epsilon_i -\icomp \eta }
+ \sum_a^\mathrm{virt}
     \frac{\varphi_a(\br)   \varphi_a(\br^\prime)}
     {\omega - \epsilon_a +\icomp \eta }
\end{equation}
expressed in terms of HF orbitals $\varphi$ and HF eigenvalues $\epsilon$. The indices $i,j,k, \dots$ denote occupied and $a,b,c, \dots$ unoccupied (or virtual) states. $\eta$ is an infinitesimal positive real number. Since HF is diagrammatic, arbitrariness in the choice of $G^{(0)}$ is avoided. Moreover, for small molecules, HF orbitals are known to be close to true Dyson orbitals. \cite{DiazTinoco2019}
For the kernel we choose the expression (Figure~\ref{fig:diagrams}c)
\begin{equation}
\label{static-vertex}
    I(1,2,3,4) = \delta(1,3)\delta(2,4)v(1,4) - \delta(1,4)\delta(3,2)W_0(1,3) \;.
\end{equation}
For $W_0 = 0$, the GWA is recovered and with $W_0 = v$, one obtains the time-dependent HF (TDHF) self-energy. \cite{Vacondio2024} Another possibility is to set $W_0 = W(\omega=0)$ where $W$ denotes the screened Coulomb interaction calculated within the random-phase approximation (RPA). $L$ then turns into the usual BSE implemented in many electronic structure codes, with the important difference that it is constructed with HF eigenvalues instead of the $GW$ ones. By choosing a static approximation to $I$, only the electron-hole part of $L$ will contribute to $\Sigma$. Eq.~\eqref{BSE} turns into a function of a single frequency which can be solved exactly by diagonalization in the particle-hole representation. \cite{Romaniello2009b} This part of $L$ is: \cite{Rohlfing2000} 
\begin{equation}
  \label{l_realspace}
  \begin{aligned}
  & L(\br_5,\br_3,\br_6,\br_3, \omega) \\
    & = -i \sum_S 
    \left[\frac{\chi_S(\br_5,\br_6)\chi^*_S(\br_3,\br_3)}{\omega - \Omega_S + i\eta} - \frac{\chi_S(\br_3,\br_3)\chi^*_S(\br_6,\br_5)}{{\omega + \Omega_S - i\eta}}\right] 
  \end{aligned}
\end{equation}
where $\Omega_S$ are the neutral excitation energies of the system and the amplitudes
\begin{equation}
\chi_S(\br_1,\br_2) = 
\sum_{ia}X^S_{ia}\varphi_a(\br_1)\varphi^*_i(\br_2) + 
\sum_{ia}Y^S_{ia}\varphi_i(\br_1)\varphi^*_a(\br_2) \;,
\end{equation}
are expressed in terms of resonant and anti-resonant transition matrix elements $X$ and $Y$.
We use the consistent notations for $X$ and $Y$ as those used in the usual Casida's equations solution.
\cite{ullrich_book}
The correlation part of the self-energy can now be written as $\Sigma = \Sigma^o + \Sigma^v$ with the contributions
\begin{equation}
\begin{aligned}
\label{sigmao}
\Sigma_{pq}^o(\omega)
 = & \sum_S \sum_k \frac{1}{\omega - \epsilon_k + \Omega_S - i \eta} \\
  & \times
 \left[\sum_{ia}  2( a i | v | q k ) ( X_{ia}^S  + Y_{ia}^S ) -
 ( k a| W_0 | q i ) X_{ia}^S 
               - ( k i | W_0 | q a ) Y_{ia}^S
  \right] \\
  & \times
   \left[ \sum_{jb} ( b j | v | p k ) ( X_{jb}^S  + Y_{jb}^S )
  \right]
  \end{aligned}
\end{equation}
and
\begin{equation}
\label{sigmav}
\begin{aligned}
\Sigma_{pq}^v(\omega)
 = & \sum_S \sum_c \frac{1}{\omega - \epsilon_c - \Omega_S
    + i \eta} \\
 & \times
 \left[\sum_{ia} 2( a i | v | q c ) ( X_{ia}^S  + Y_{ia}^S )
 -( c i | W_0 | q a ) X_{ia}^S 
                - ( c a | W_0 | q i ) Y_{ia}^S
  \right] \\
  & \times
   \left[ \sum_{jb} ( b j | v | p c  ) ( X_{jb}^S  + Y_{jb}^S )
  \right]
  \end{aligned}
\end{equation}
The factor of $2$ comes from spin-summation and is absent in the exchange terms. The four-center integrals for $v$ and $W_0$ are defined with the chemists' notation:
\begin{equation}
\label{eq:eri}
( pq | v | rs ) = \int d\br \int d\br'
   \varphi^*_p(\br) \varphi_q(\br) v(\br,\br')
   \varphi^*_r(\br') \varphi_s(\br') \; .
\end{equation}
For a detailed derivation, we refer to the SI. Other authors have already used this scheme presented there with the TDHF kernel \cite{Maggio2017, Maggio2018, Vlcek2019, Vacondio2024} and we extend it here by using a screened exchange kernel.
Based on the prior knowledge about BSE success and the TDHF mixed performance to describe
neutral excitations, \cite{blase_jpcl2020} we expect this improvement to be significant.
The Dyson-like structure of the equations ensure that the kernel is consistently included to infinite-order in $L$ and hence in $\Sigma$. 
It adds diagrams to the self-energy which describe electron-hole interactions and are important at short inter-electronic distances. \cite{Irmler2019a}
We go beyond approaches that include the vertex in $\Sigma$ to first order only,\cite{Bobbert1994, Ummels1998, Schindlmayr1998} leading for instance to $G3W2$ vertex-corrections \cite{Kuwahara2016, Kutepov2016, Kutepov2017a, Bruneval2024} and approximations like its completely statically screened version,\cite{Gruneis2014, Forster2022} SOSEX,\cite{Ren2015, Loos2018, Vacondio2022} or subsets of $G3W2$.\cite{Stefanucci2014, Pavlyukh2016, Pavlyukh2020} Using different kernels in $L$ and $\Sigma$ is possible, but we show here that the kernels should be kept consistent.
The different approximations used in this work are summarized in Table~\ref{tab:approx}.

\begin{table}[t!]
\centering
\resizebox{\textwidth}{!}{
\begin{tabular}{lllll}
\toprule
$I(6,7,5,8)$ in $L$ & $L$ & 
$I(3,5,4,6)$ in $\Sigma$ & $\Sigma$
& Vernacular name \\
\midrule
0 & $L^{(0)}$ &
$v(3,6) \delta(3,4) \delta(5,6)$ 
- $v(3,4) \delta(3,6) \delta(4,5)$ &
$\Sigma^\mathrm{PT2}$ &
PT2, GF2, or 2-Born \\

 $v(6,8) \delta(6,5) \delta(7,8)$  &
$L^\mathrm{TDH}$  &
$v(3,6) \delta(3,4) \delta(5,6)$ &
$GW@L^\mathrm{TDH}$ &
 standard $GW$ \\
 
$v(6,8) \delta(6,5) \delta(7,8)$
- $v(6,5) \delta(6,8) \delta(7,5)$ &
$L^\mathrm{TDHF}$  &
$v(3,6) \delta(3,4) \delta(5,6)$ &
$GW@L^\mathrm{TDHF}$ &
 $GW$ with TDHF screening \\

$v(6,8) \delta(6,5) \delta(7,8)$
- $W_0(6,5) \delta(6,8) \delta(7,5)$ &
$L^\mathrm{BSE}$  &
$v(3,6) \delta(3,4) \delta(5,6)$ &
$GW@L^\mathrm{BSE}$ &
 $GW$ with BSE screening \\

$v(6,8) \delta(6,5) \delta(7,8)$
- $v(6,5) \delta(6,8) \delta(7,5)$ &
$L^\mathrm{TDHF}$  &
$v(3,6) \delta(3,4) \delta(5,6)$ 
- $v(3,4) \delta(3,6) \delta(4,5)$ &
$\Sigma^\mathrm{TDHF}$ &
TDHF self-energy \\

$v(6,8) \delta(6,5) \delta(7,8)$
- $W_0(6,5) \delta(6,8) \delta(7,5)$ &
$L^\mathrm{BSE}$  &
$v(3,6) \delta(3,4) \delta(5,6)$ 
- $W_0(3,4) \delta(3,6) \delta(4,5)$ &
$\Sigma^\mathrm{BSE}$ &
BSE self-energy \\
\bottomrule
\end{tabular}}
\caption{
Summary of the different infinite order approximations used in this work.}
\label{tab:approx}
\end{table}

In the following, we discuss numerical results. We first test different vertex corrected schemes on the GW100 test set of first molecular ionization potentials.\cite{VanSetten2015} We perform all calculations with MOLGW \cite{Bruneval2016a} and BAND \cite{TeVelde1991, Spadetto2023} using the def2-qzvpp basis set and use the corresponding CCSD(T) values from
Ref.~\citenum{Monzel2023} as reference. 

\begin{figure*}[hbt!]
  \includegraphics[width=\linewidth]{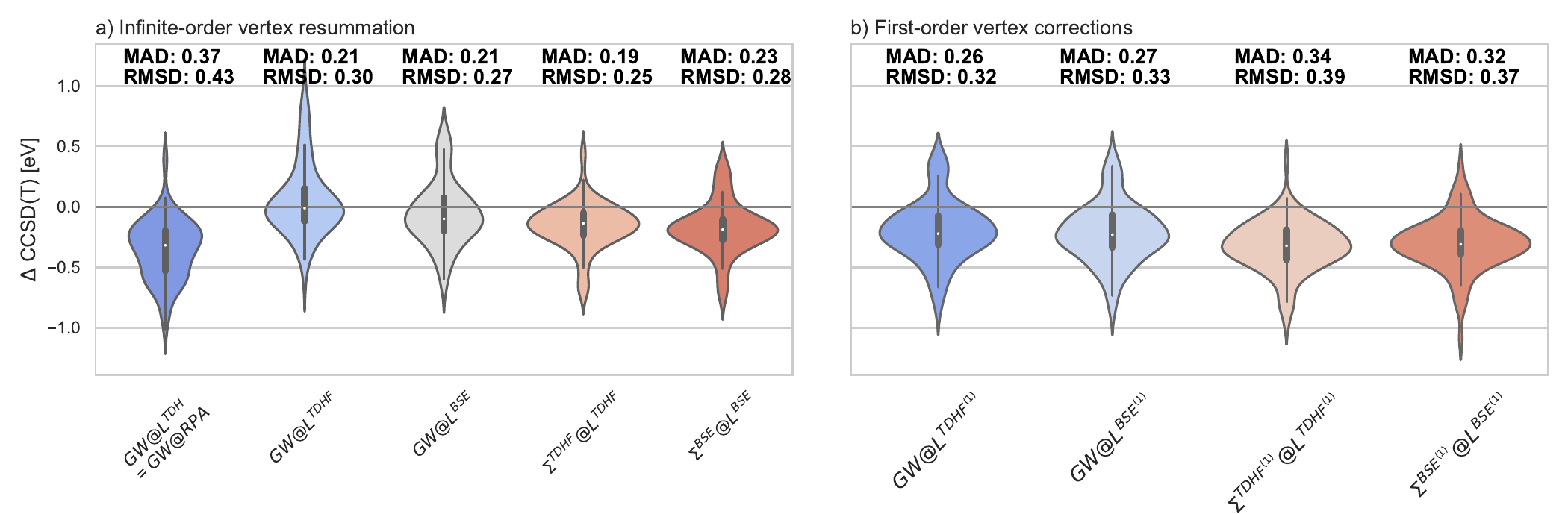}
  \caption{Errors of $GW$ and several vertex corrected schemes with respect to CCSD(T) in eV of the HOMO of the molecules contained in the GW100 set 
  for infinite vertex resummation (panel a) and first-order only (panel b).}
  \label{fig:gw100violins}
\end{figure*}

\begin{figure*}[hbt!]
  \includegraphics[width=\linewidth]{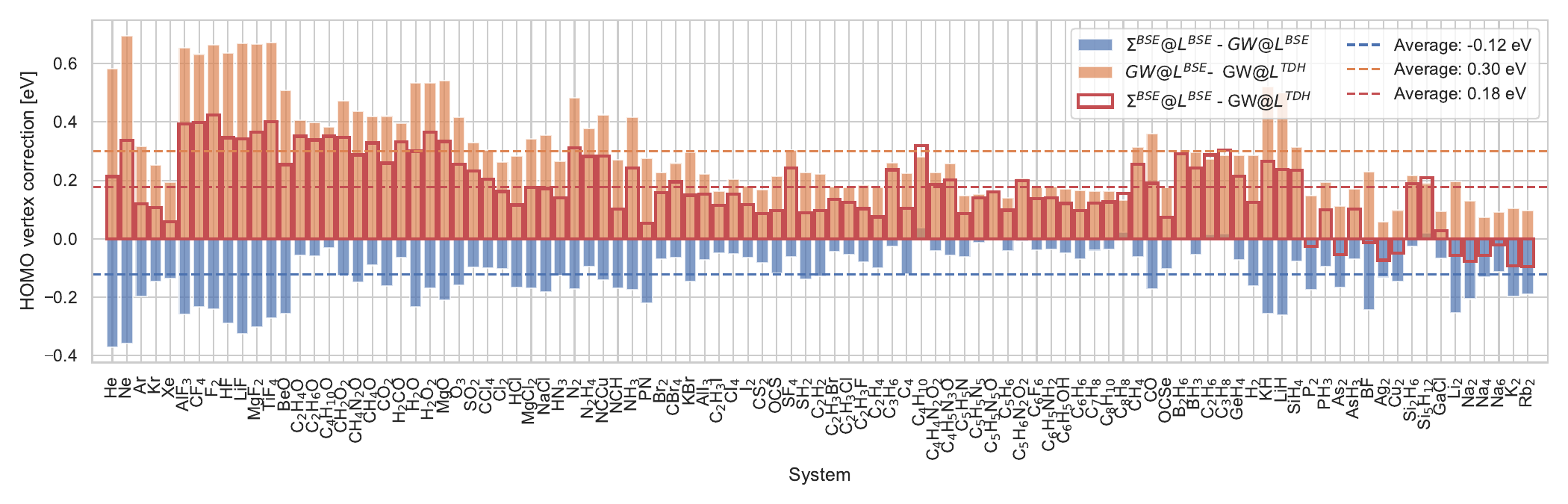}
  \caption{Vertex corrections in eV of the highest occupied molecular orbital of the molecules in the GW100 set. Besides the rare gases, the molecules are sorted by decreasing electronegativity of the element most represented in the HOMO.}
  \label{fig:gw100vertexcorr}
\end{figure*}

Figure~\ref{fig:gw100violins} shows their error distributions of several vertex-corrected schemes compared to CCSD(T) together with mean absolute deviations (MAD) and root mean square deviations (RMSD). The leftmost violin in Fig.~\ref{fig:gw100violins}a) shows the errors of $GW$@RPA. The next two violins show the errors for the $GW$ self-energy with $L$ calculated with TDHF and BSE, respectively. The final two plots show the results for $\Sigma^{TDHF}@L^{TDHF}$ and $\Sigma^{BSE}@L^{BSE}$ which both include the vertex in Eq.~\eqref{static-vertex} consistently to infinite order in $L$ and $\Sigma$. All four vertex-corrected schemes give major improvements over $GW$. This is also true for $GW@L^{TDHF}$ and $GW@L^{BSE}$ which only include the vertex in $L$.
Therefore, our results temper the strong conclusions
of Lewis \textit{et. al.}, \cite{Lewis2019}
when they claim that the efforts to improve
the screening part of the self-energy are worsening the results.
However, $GW@L^{TDHF}$ and $GW@L^{BSE}$ (to a lesser extent) lead to major errors for some molecules.
The kernel in $\Sigma$ balances the sizable effect of the kernel in $L$, leading to consistent improvements over $GW$@RPA.

To understand this behavior, we show that the vertex corrections systematically have opposite signs in Fig.~\ref{fig:gw100vertexcorr} and therefore partially and sometimes completely cancel.
The orange bars show the magnitude of the vertex correction beyond TDH in $L$ (corresponding to the third violin in Fig.~\ref{fig:gw100violins}a)), and the blue bars the magnitude of the vertex correction in $\Sigma$ beyond $GW$ (corresponding to the last violin in Fig.~\ref{fig:gw100violins}a)).
The red boxes show the difference between $GW$@RPA and $\Sigma^{BSE}@L^{BSE}$, which is the sum of the blue and orange bars. The BSE kernel describes the electron-hole interaction missing in $GW$@RPA which stabilizes the cation and therefore lowers the HOMO energy. In some cases exceeding 0.6 eV, this effect is sizable for most molecules in GW100 and frequently the HOMO energy is overcorrected. The vertex correction in $\Sigma$ has the opposite effect and reduces the HOMO energy further. The effect of the vertex is generally stronger for $L$ than for $\Sigma$ and therefore the combination of both vertex corrections lowers the HOMO. Both vertices combined lead to the observed improvement of $\Sigma^{BSE}@L^{BSE}$ over $GW$@RPA.

As shown in the SI (Fig.~S1), a similar picture is obtained for the TDHF
screening and self-energy approximations. Our results qualitatively agree with Ref.~\citenum{Vacondio2024}.
With average values of 0.41 and -0.21 eV, the effect of the individual vertex corrections in $L$ and $\Sigma$ is significantly larger. However, with 0.2 eV on average, the combined effect of the vertex correction is comparable to BSE. The BSE vertex correction accounts for higher-order vertex diagrams not included in $\Sigma^{TDHF}@L^{TDHF}$. The smaller magnitudes of the vertex corrections in $L$ and $\Sigma$ with the BSE vertex indicate further cancellations between these higher-order diagrams.

Further insight into the cancellation of vertex corrections is provided in Fig.~\ref{fig:gw100violins}b) where we show the errors of the same vertex-corrected schemes as in Fig.~\ref{fig:gw100violins}a), but in all cases truncated to first order. For the polarizability, this means that RPA screening is modified by including only one diagram of first order in $W_0$. \cite{Bechstedt1997, Schindlmayr1998, Kuwahara2016} Including the same vertex diagram in $\Sigma$ the SOSEX self-energy is obtained with the bare vertex,\cite{Bruneval2024} and the screened vertex leads to a second-order term similar to SOSEX but with the bare $v$ replaced by the statically screened one. In this scheme, the vertex correction is consistent in $L$ and $\Sigma$ since the next-to-leading order diagram is added to both quantities. We refer to the SI (section~S3) for detailed derivations.

Including the kernel to first order only has generally a much smaller effect than the infinite-order resummations (0.17~eV on average vs. 0.41~eV for $L$ and -0.13 vs. -0.21~eV for $\Sigma$) As shown in the Supporting Information Fig.~S2 and Fig.~S3, adding the same vertex correction to $W$ and $\Sigma$ results in HOMO energies almost indistinguishable from $GW$@RPA. The same conclusion has already been drawn based on results for the band gap of silicon \cite{Bobbert1994} and for a one-dimensional semiconductor \cite{DeGroot1996} and we confirm here its validity for molecules. This almost complete cancellation of the next-to-leading order terms in $L$ and $\Sigma$ rationalizes the good performance of $GW$@RPA for calculating QP energies.

\begin{figure}[hbt!]
  \includegraphics[width=0.5\linewidth]{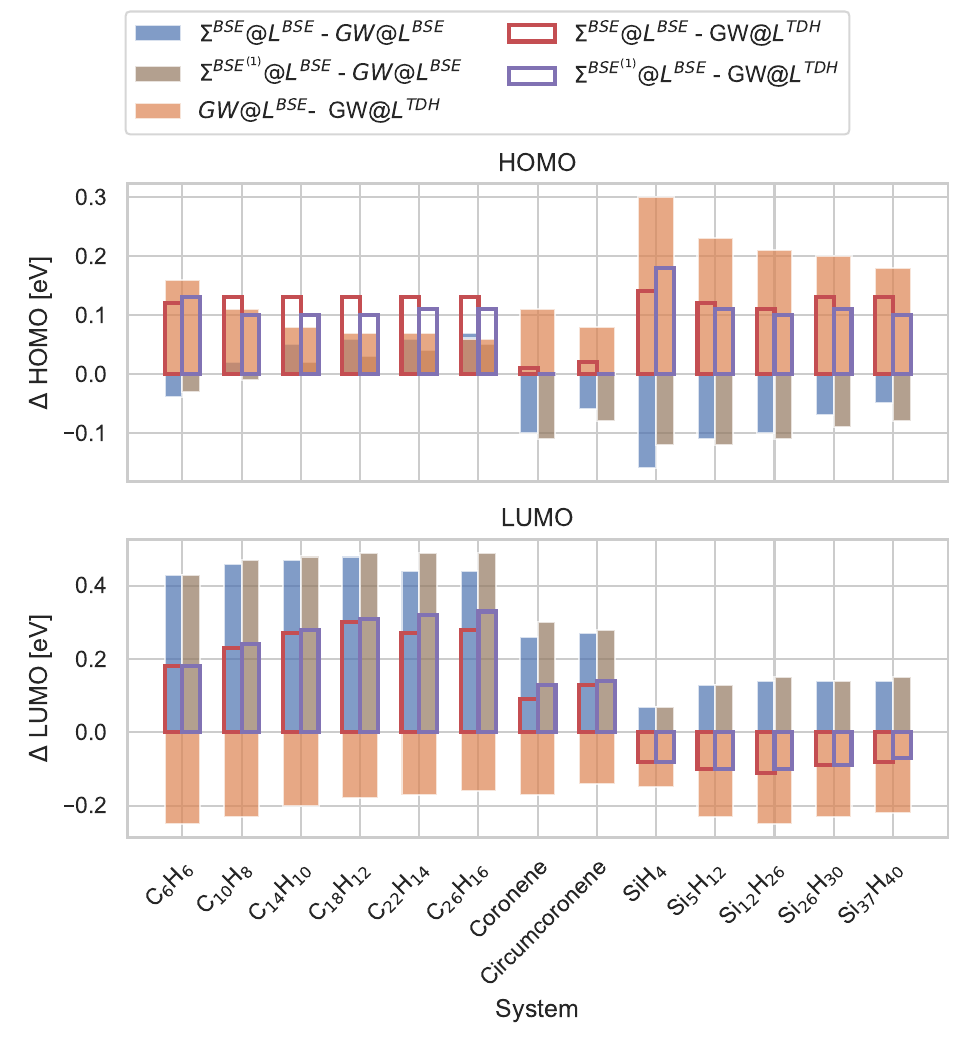}
  \caption{Screened exchange vertex corrections in eV of the HOMO (top), LUMO (bottom) for linear acenes, non-linear acenes, and passivated silicon clusters of increasing size.}
  \label{fig:ClustersVertexCorr}
\end{figure}

\begin{figure}[hbt!]
  \includegraphics[width=0.5\linewidth]{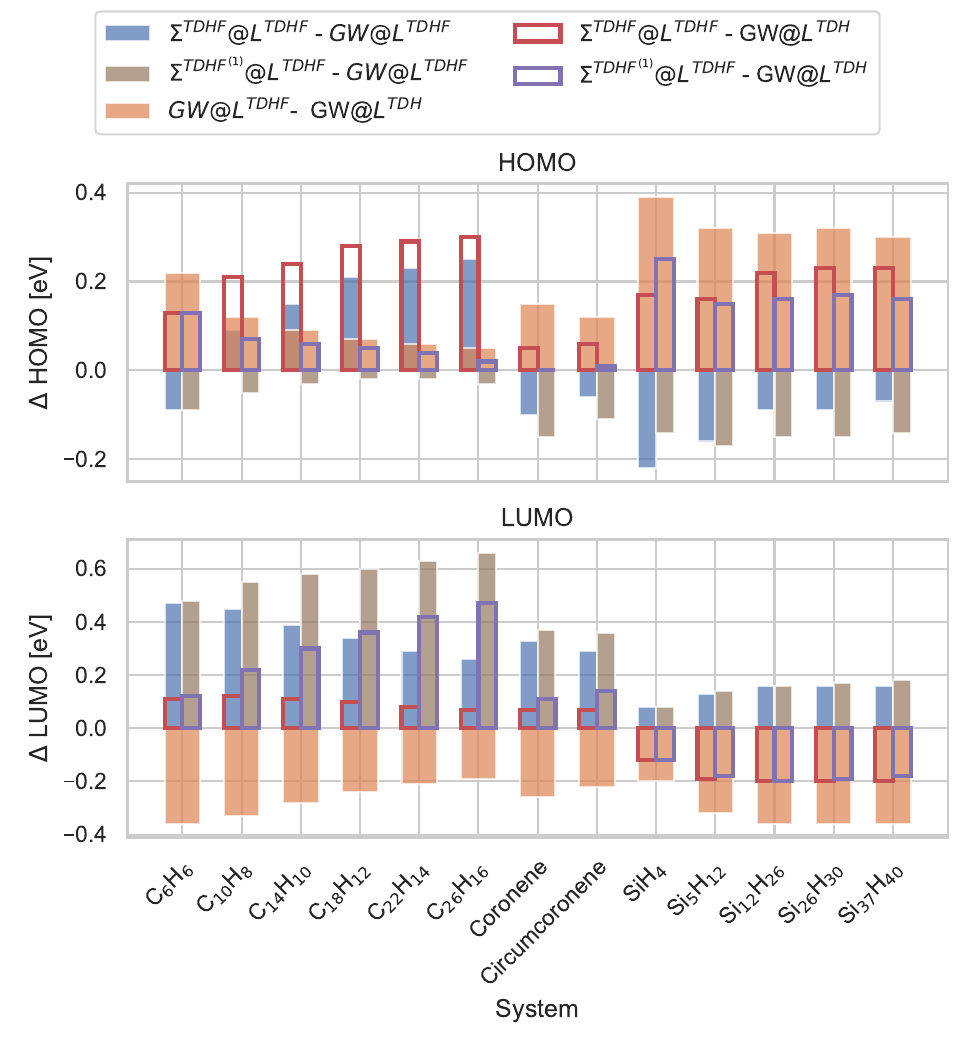}
  \caption{Bare exchange vertex corrections in eV of the HOMO (top) and LUMO (bottom) for linear acenes, non-linear acenes, and passivated silicon clusters of increasing size.}
  \label{fig:ClustersVertexCorrTDHF}
\end{figure}

Finally, in Fig.~\ref{fig:ClustersVertexCorr}, we show the magnitude of the different BSE vertex corrections for the HOMO and LUMO energies of molecules of systematically increasing size: Linear acenes ranging from a single benzene ring (C$_6$H$_6$) up to hexacene (C$_{26}$H$_{16}$), coronene and circumcoronene, as well as passivated silicon clusters with up to 37 silicon atoms. 
For the BSE vertex, we find the magnitude of the vertex correction to be almost independent of the system size and with about 0.1 eV to be rather small. The effect on the LUMO is with 0.2 eV about twice as large. This observation is consistent with Ref.~\citenum{Vlcek2019}. While initially increasing, the magnitude of the vertex correction stays approximately constant for the linear acenes and silicon clusters. We also notice that the first-order truncation of $\Sigma^{BSE}$ is always a good approximation.

Figure~\ref{fig:ClustersVertexCorrTDHF} shows the same information for the TDHF vertex. As for GW100, the magnitudes of the individual vertex corrections in $\Sigma$ and $L$ are larger than for the BSE vertex. Also, the total vertex correction is much larger than for the BSE vertex. Moreover, we observe especially for the linear acenes that the infinite-order resummation of the TDHF vertex in $\Sigma^{TDHF}$ leads to a rapidly increasing vertex correction for the HOMO. At the same time, its first-order approximation (SOSEX) goes to almost zero. The opposite can be observed for the LUMO. This inconsistency indicates the importance of screened vertices for larger systems.

In conclusion, several vertex-corrected schemes have been investigated over the last decades to improve over the simple GWA for QP energies. Cancellations between vertex corrections have been observed early on for simple (model) systems. \cite{Mahan1989, delSole1994, DeGroot1996} Despite immense implications for practical $GW$ calculations, these results have never been generally confirmed using realistic, non-local vertices. 

With this work, we have filled this gap. To rationalize the success of the GWA for calculating QP energies, we have investigated several vertex corrections beyond the GWA. We benchmarked these methods for systems ranging from small and medium molecules in the GW100 set, over linear and non-linear acenes, to silicon clusters. We have used the TDHF vertex as obtained from the HF self-energy, which adds infinite-order particle-hole diagrams to $L$ and $\Sigma$ as well as a BSE vertex which statically screens these diagrams. Especially for larger molecules it becomes decisive to use a screened vertex correction. 

By restricting infinite-order vertex summation to first-order only, we have also performed calculations that only include the next-to-leading order correction to $L$ and $\Sigma$. Both corrections effectively cancel for HOMO QP energies, suggesting an order-by-order expansion of $L$ and $\Sigma$ beyond $GW$@RPA to be inefficient. Despite being of low order in perturbation theory, it accounts for the most important signatures of electron correlation for a charged excitation.

We have rationalized why schemes that add a vertex correction to either 
the response function or the self-energy have been unsuccessful. \cite{Lewis2019, Bruneval2024} The cancellations between these vertices are far-reaching and they both must be included to obtain systematic improvements over $GW$. To improve over $GW$@RPA, infinite-order resummations of the vertex function are needed in both $L$ and $\Sigma$. Moving forward, dynamical vertex corrections could be explored. These would allow for the inclusion of the yet missing particle-particle channel to the self-energy which is important in the strongly correlated regime. \cite{Gunnarsson2016, Mejuto-Zaera2022a}

\begin{acknowledgement}
Part of this work was performed using HPC resources from GENCI–TGCC (Grant 2024-gen6018). 
AF acknowledges the use of supercomputer facilities at SURFsara sponsored by NWO Physical Sciences, with financial support from The Netherlands Organization for Scientific Research (NWO).
\end{acknowledgement}

\begin{suppinfo}
Detailed derivations and discussions of all equations, additional computational details, and all QP energies calculated in this work,
\end{suppinfo}

\begin{tocentry}
    \includegraphics[width=\linewidth]{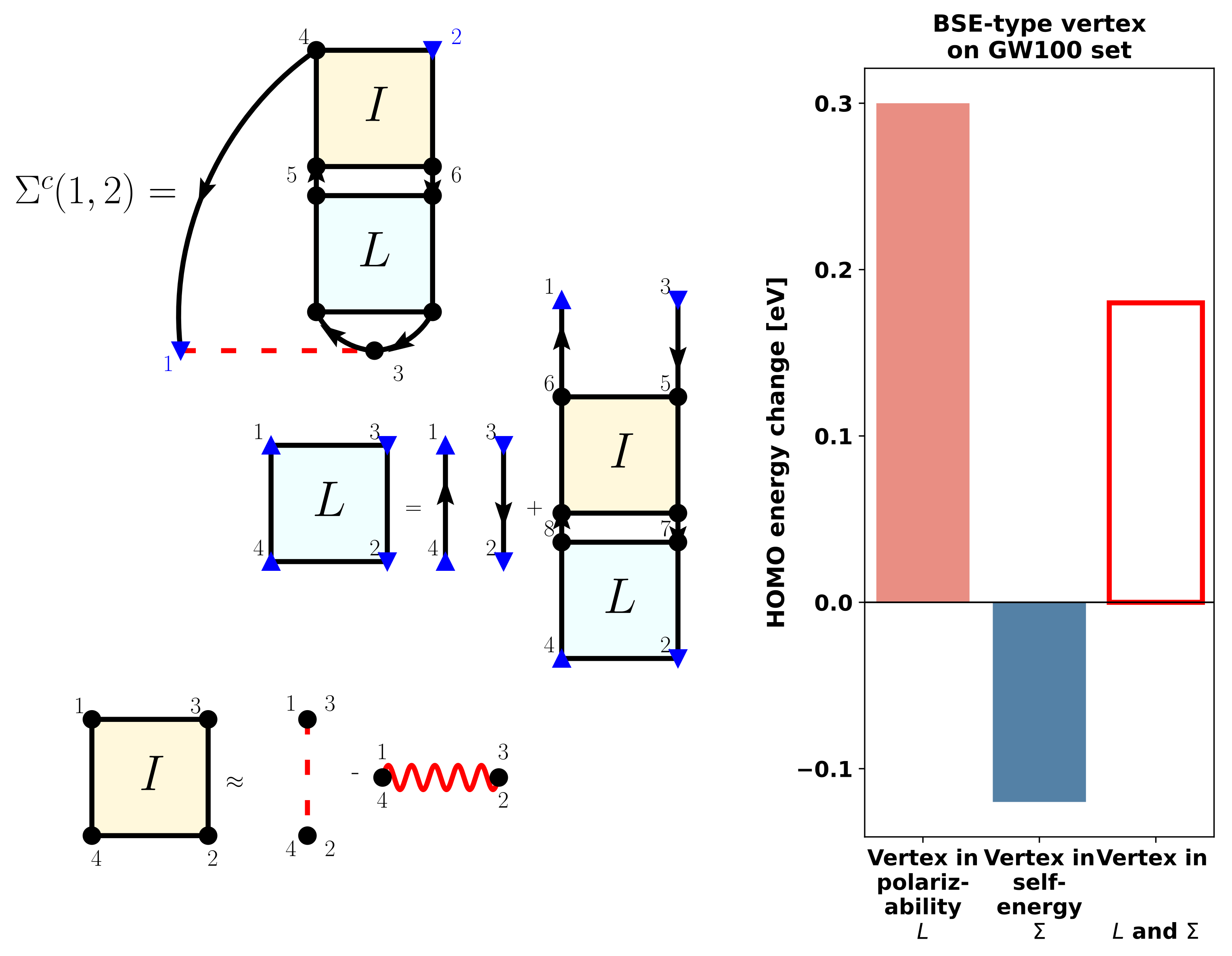}
\end{tocentry}

\bibliography{all.bib,biblio_fabien.bib}% Produces the bibliography via BibTeX.

\providecommand{\latin}[1]{#1}
\makeatletter
\providecommand{\doi}
  {\begingroup\let\do\@makeother\dospecials
  \catcode`\{=1 \catcode`\}=2 \doi@aux}
\providecommand{\doi@aux}[1]{\endgroup\texttt{#1}}
\makeatother
\providecommand*\mcitethebibliography{\thebibliography}
\csname @ifundefined\endcsname{endmcitethebibliography}  {\let\endmcitethebibliography\endthebibliography}{}
\begin{mcitethebibliography}{118}
\providecommand*\natexlab[1]{#1}
\providecommand*\mciteSetBstSublistMode[1]{}
\providecommand*\mciteSetBstMaxWidthForm[2]{}
\providecommand*\mciteBstWouldAddEndPuncttrue
  {\def\EndOfBibitem{\unskip.}}
\providecommand*\mciteBstWouldAddEndPunctfalse
  {\let\EndOfBibitem\relax}
\providecommand*\mciteSetBstMidEndSepPunct[3]{}
\providecommand*\mciteSetBstSublistLabelBeginEnd[3]{}
\providecommand*\EndOfBibitem{}
\mciteSetBstSublistMode{f}
\mciteSetBstMaxWidthForm{subitem}{(\alph{mcitesubitemcount})}
\mciteSetBstSublistLabelBeginEnd
  {\mcitemaxwidthsubitemform\space}
  {\relax}
  {\relax}

\bibitem[Hedin(1965)]{Hedin1965}
Hedin,~L. {New method for calculating the one-particle Green's function with application to the electron-gas problem}. \emph{Phys. Rev.} \textbf{1965}, \emph{139}, A796\relax
\mciteBstWouldAddEndPuncttrue
\mciteSetBstMidEndSepPunct{\mcitedefaultmidpunct}
{\mcitedefaultendpunct}{\mcitedefaultseppunct}\relax
\EndOfBibitem
\bibitem[Aryasetiawan and Gunnarsson(1998)Aryasetiawan, and Gunnarsson]{Aryasetiawan1998}
Aryasetiawan,~F.; Gunnarsson,~O. {The GW method}. \emph{Reports Prog. Phys.} \textbf{1998}, \emph{61}, 237--312\relax
\mciteBstWouldAddEndPuncttrue
\mciteSetBstMidEndSepPunct{\mcitedefaultmidpunct}
{\mcitedefaultendpunct}{\mcitedefaultseppunct}\relax
\EndOfBibitem
\bibitem[Golze \latin{et~al.}(2019)Golze, Dvorak, and Rinke]{Golze2019}
Golze,~D.; Dvorak,~M.; Rinke,~P. {The GW Compendium: A Practical Guide to Theoretical Photoemission Spectroscopy}. \emph{Front. Chem.} \textbf{2019}, \emph{7}, 377\relax
\mciteBstWouldAddEndPuncttrue
\mciteSetBstMidEndSepPunct{\mcitedefaultmidpunct}
{\mcitedefaultendpunct}{\mcitedefaultseppunct}\relax
\EndOfBibitem
\bibitem[Lundqvist(1967)]{lundqvist_pkm1967}
Lundqvist,~B. Single particle spectrum of the degenerate electron gas. \emph{Physik der Kondensierten Materie} \textbf{1967}, \emph{6}, 193--205\relax
\mciteBstWouldAddEndPuncttrue
\mciteSetBstMidEndSepPunct{\mcitedefaultmidpunct}
{\mcitedefaultendpunct}{\mcitedefaultseppunct}\relax
\EndOfBibitem
\bibitem[Strinati \latin{et~al.}(1980)Strinati, Mattausch, and Hanke]{strinati_prl1980}
Strinati,~G.; Mattausch,~H.~J.; Hanke,~W. Dynamical Correlation Effects on the Quasiparticle Bloch States of a Covalent Crystal. \emph{Phys. Rev. Lett.} \textbf{1980}, \emph{45}, 290--294\relax
\mciteBstWouldAddEndPuncttrue
\mciteSetBstMidEndSepPunct{\mcitedefaultmidpunct}
{\mcitedefaultendpunct}{\mcitedefaultseppunct}\relax
\EndOfBibitem
\bibitem[Strinati \latin{et~al.}(1982)Strinati, Mattausch, and Hanke]{Strinati1982a}
Strinati,~G.; Mattausch,~H.~J.; Hanke,~W. {Dynamical aspects of correlation corrections in a covalent crystal}. \emph{Phys. Rev. B} \textbf{1982}, \emph{25}, 2867--2888\relax
\mciteBstWouldAddEndPuncttrue
\mciteSetBstMidEndSepPunct{\mcitedefaultmidpunct}
{\mcitedefaultendpunct}{\mcitedefaultseppunct}\relax
\EndOfBibitem
\bibitem[Hybertsen and Louie(1985)Hybertsen, and Louie]{Hybertsen1985}
Hybertsen,~M.~S.; Louie,~S.~G. {First-principles theory of quasiparticles: Calculation of band gaps in semiconductors and insulators}. \emph{Phys. Rev. Lett.} \textbf{1985}, \emph{55}, 1418--1421\relax
\mciteBstWouldAddEndPuncttrue
\mciteSetBstMidEndSepPunct{\mcitedefaultmidpunct}
{\mcitedefaultendpunct}{\mcitedefaultseppunct}\relax
\EndOfBibitem
\bibitem[Hybertsen and Louie(1986)Hybertsen, and Louie]{Hybertsen1986}
Hybertsen,~M.~S.; Louie,~S.~G. {Electron correlation in semiconductors and insulators: Band gaps and quasiparticle energies}. \emph{Phys. Rev. B} \textbf{1986}, \emph{34}, 5390\relax
\mciteBstWouldAddEndPuncttrue
\mciteSetBstMidEndSepPunct{\mcitedefaultmidpunct}
{\mcitedefaultendpunct}{\mcitedefaultseppunct}\relax
\EndOfBibitem
\bibitem[Godby \latin{et~al.}(1988)Godby, Schl{\"{u}}ter, and Sham]{Godby1988}
Godby,~R.~W.; Schl{\"{u}}ter,~M.; Sham,~L.~J. {Self-energy operators and exchange-correlation potentials in semiconductors}. \emph{Phys. Rev. B} \textbf{1988}, \emph{37}, 10159--10175\relax
\mciteBstWouldAddEndPuncttrue
\mciteSetBstMidEndSepPunct{\mcitedefaultmidpunct}
{\mcitedefaultendpunct}{\mcitedefaultseppunct}\relax
\EndOfBibitem
\bibitem[Sch{\"{o}}ne and Eguiluz(1998)Sch{\"{o}}ne, and Eguiluz]{Schone1998}
Sch{\"{o}}ne,~W.~D.; Eguiluz,~A.~G. {Self-consistent calculations of quasiparticle states in metals and semiconductors}. \emph{Phys. Rev. Lett.} \textbf{1998}, \emph{81}, 1662--1665\relax
\mciteBstWouldAddEndPuncttrue
\mciteSetBstMidEndSepPunct{\mcitedefaultmidpunct}
{\mcitedefaultendpunct}{\mcitedefaultseppunct}\relax
\EndOfBibitem
\bibitem[Onida \latin{et~al.}(1995)Onida, Reining, Godby, {Del Sole}, and Andreoni]{Onida1995}
Onida,~G.; Reining,~L.; Godby,~R.~W.; {Del Sole},~R.; Andreoni,~W. {Ab initio calculations of the quasiparticle and absorption spectra of clusters: The sodium tetramer}. \emph{Phys. Rev. Lett.} \textbf{1995}, \emph{75}, 818--821\relax
\mciteBstWouldAddEndPuncttrue
\mciteSetBstMidEndSepPunct{\mcitedefaultmidpunct}
{\mcitedefaultendpunct}{\mcitedefaultseppunct}\relax
\EndOfBibitem
\bibitem[Ishii \latin{et~al.}(2001)Ishii, Ohno, Kawazoe, and Louie]{Ishii2001}
Ishii,~S.; Ohno,~K.; Kawazoe,~Y.; Louie,~S.~G. {Ab initio GW quasiparticle energies of small sodium clusters by an all-electron mixed-basis approach}. \emph{Phys. Rev. B - Condens. Matter Mater. Phys.} \textbf{2001}, \emph{63}, 155104\relax
\mciteBstWouldAddEndPuncttrue
\mciteSetBstMidEndSepPunct{\mcitedefaultmidpunct}
{\mcitedefaultendpunct}{\mcitedefaultseppunct}\relax
\EndOfBibitem
\bibitem[Ishii \latin{et~al.}(2002)Ishii, Ohno, Kawazoe, and Louie]{Ishii2002}
Ishii,~S.; Ohno,~K.; Kawazoe,~Y.; Louie,~S.~G. {Ab initio GW quasiparticle calculation of small alkali-metal clusters}. \emph{Phys. Rev. B - Condens. Matter Mater. Phys.} \textbf{2002}, \emph{65}, 245109\relax
\mciteBstWouldAddEndPuncttrue
\mciteSetBstMidEndSepPunct{\mcitedefaultmidpunct}
{\mcitedefaultendpunct}{\mcitedefaultseppunct}\relax
\EndOfBibitem
\bibitem[Rohlfing(2000)]{Rohlfing2000a}
Rohlfing,~M. {Excited states of molecules from Green's function perturbation techniques}. \emph{Int. J. Quantum Chem.} \textbf{2000}, \emph{80}, 807--815\relax
\mciteBstWouldAddEndPuncttrue
\mciteSetBstMidEndSepPunct{\mcitedefaultmidpunct}
{\mcitedefaultendpunct}{\mcitedefaultseppunct}\relax
\EndOfBibitem
\bibitem[Grossman \latin{et~al.}(2001)Grossman, Rohlfing, Mitas, Louie, and Cohen]{Grossman2001}
Grossman,~J.~C.; Rohlfing,~M.; Mitas,~L.; Louie,~S.~G.; Cohen,~M.~L. {High accuracy many-body calculational approaches for excitations in molecules}. \emph{Phys. Rev. Lett.} \textbf{2001}, \emph{86}, 472--475\relax
\mciteBstWouldAddEndPuncttrue
\mciteSetBstMidEndSepPunct{\mcitedefaultmidpunct}
{\mcitedefaultendpunct}{\mcitedefaultseppunct}\relax
\EndOfBibitem
\bibitem[Tiago and Chelikowsky(2006)Tiago, and Chelikowsky]{Tiago2006}
Tiago,~M.~L.; Chelikowsky,~J.~R. {Optical excitations in organic molecules, clusters, and defects studied by first-principles Green's function methods}. \emph{Phys. Rev. B} \textbf{2006}, \emph{73}, 205334\relax
\mciteBstWouldAddEndPuncttrue
\mciteSetBstMidEndSepPunct{\mcitedefaultmidpunct}
{\mcitedefaultendpunct}{\mcitedefaultseppunct}\relax
\EndOfBibitem
\bibitem[Rostgaard \latin{et~al.}(2010)Rostgaard, Jacobsen, and Thygesen]{Rostgaard2010}
Rostgaard,~C.; Jacobsen,~K.~W.; Thygesen,~K.~S. {Fully self-consistent GW calculations for molecules}. \emph{Phys. Rev. B} \textbf{2010}, \emph{81}, 085103\relax
\mciteBstWouldAddEndPuncttrue
\mciteSetBstMidEndSepPunct{\mcitedefaultmidpunct}
{\mcitedefaultendpunct}{\mcitedefaultseppunct}\relax
\EndOfBibitem
\bibitem[Blase \latin{et~al.}(2011)Blase, Attaccalite, and Olevano]{Blase2011}
Blase,~X.; Attaccalite,~C.; Olevano,~V. {First-principles GW calculations for fullerenes, porphyrins, phtalocyanine, and other molecules of interest for organic photovoltaic applications}. \emph{Phys. Rev. B} \textbf{2011}, \emph{83}, 115103\relax
\mciteBstWouldAddEndPuncttrue
\mciteSetBstMidEndSepPunct{\mcitedefaultmidpunct}
{\mcitedefaultendpunct}{\mcitedefaultseppunct}\relax
\EndOfBibitem
\bibitem[Ke(2011)]{Ke2011}
Ke,~S.~H. {All-electron GW methods implemented in molecular orbital space: Ionization energy and electron affinity of conjugated molecules}. \emph{Phys. Rev. B} \textbf{2011}, \emph{84}, 205415\relax
\mciteBstWouldAddEndPuncttrue
\mciteSetBstMidEndSepPunct{\mcitedefaultmidpunct}
{\mcitedefaultendpunct}{\mcitedefaultseppunct}\relax
\EndOfBibitem
\bibitem[Bruneval(2012)]{Bruneval2012}
Bruneval,~F. {Ionization energy of atoms obtained from GW self-energy or from random phase approximation total energies}. \emph{J. Chem. Phys.} \textbf{2012}, \emph{136}, 194107\relax
\mciteBstWouldAddEndPuncttrue
\mciteSetBstMidEndSepPunct{\mcitedefaultmidpunct}
{\mcitedefaultendpunct}{\mcitedefaultseppunct}\relax
\EndOfBibitem
\bibitem[K{\"{o}}rzd{\"{o}}rfer and Marom(2012)K{\"{o}}rzd{\"{o}}rfer, and Marom]{Korzdorfer2012}
K{\"{o}}rzd{\"{o}}rfer,~T.; Marom,~N. {Strategy for finding a reliable starting point for G 0W 0 demonstrated for molecules}. \emph{Phys. Rev. B - Condens. Matter Mater. Phys.} \textbf{2012}, \emph{86}, 041110(R)\relax
\mciteBstWouldAddEndPuncttrue
\mciteSetBstMidEndSepPunct{\mcitedefaultmidpunct}
{\mcitedefaultendpunct}{\mcitedefaultseppunct}\relax
\EndOfBibitem
\bibitem[Duchemin \latin{et~al.}(2012)Duchemin, Deutsch, and Blase]{Duchemin2012}
Duchemin,~I.; Deutsch,~T.; Blase,~X. {Short-range to long-range charge-transfer excitations in the zincbacteriochlorin-bacteriochlorin complex: A bethe-salpeter study}. \emph{Phys. Rev. Lett.} \textbf{2012}, \emph{109}, 167801\relax
\mciteBstWouldAddEndPuncttrue
\mciteSetBstMidEndSepPunct{\mcitedefaultmidpunct}
{\mcitedefaultendpunct}{\mcitedefaultseppunct}\relax
\EndOfBibitem
\bibitem[F{\"{o}}rster and Visscher(2022)F{\"{o}}rster, and Visscher]{Forster2022c}
F{\"{o}}rster,~A.; Visscher,~L. {Quasiparticle Self-Consistent GW-Bethe-Salpeter equation calculations for large chromophoric systems}. \emph{J. Chem. Theory Comput.} \textbf{2022}, \emph{18}, 6779--6793\relax
\mciteBstWouldAddEndPuncttrue
\mciteSetBstMidEndSepPunct{\mcitedefaultmidpunct}
{\mcitedefaultendpunct}{\mcitedefaultseppunct}\relax
\EndOfBibitem
\bibitem[Allen \latin{et~al.}(2024)Allen, Nguyen, and Neuhauser]{Allen2024}
Allen,~T.; Nguyen,~M.; Neuhauser,~D. {GW with hybrid functionals for large molecular systems}. \emph{arXiv:2405.12306v1} \textbf{2024}, 1--18\relax
\mciteBstWouldAddEndPuncttrue
\mciteSetBstMidEndSepPunct{\mcitedefaultmidpunct}
{\mcitedefaultendpunct}{\mcitedefaultseppunct}\relax
\EndOfBibitem
\bibitem[Thygesen and Rubio(2009)Thygesen, and Rubio]{Thygesen2009}
Thygesen,~K.~S.; Rubio,~A. {Renormalization of molecular quasiparticle levels at metal-molecule interfaces: Trends across binding regimes}. \emph{Phys. Rev. Lett.} \textbf{2009}, \emph{102}, 046802\relax
\mciteBstWouldAddEndPuncttrue
\mciteSetBstMidEndSepPunct{\mcitedefaultmidpunct}
{\mcitedefaultendpunct}{\mcitedefaultseppunct}\relax
\EndOfBibitem
\bibitem[Liu \latin{et~al.}(2019)Liu, {Da Jornada}, Louie, and Neaton]{Liu2019}
Liu,~Z.~F.; {Da Jornada},~F.~H.; Louie,~S.~G.; Neaton,~J.~B. {Accelerating GW-Based Energy Level Alignment Calculations for Molecule-Metal Interfaces Using a Substrate Screening Approach}. \emph{J. Chem. Theory Comput.} \textbf{2019}, \emph{15}, 4218--4227\relax
\mciteBstWouldAddEndPuncttrue
\mciteSetBstMidEndSepPunct{\mcitedefaultmidpunct}
{\mcitedefaultendpunct}{\mcitedefaultseppunct}\relax
\EndOfBibitem
\bibitem[Adeniran and Liu(2021)Adeniran, and Liu]{Adeniran2021}
Adeniran,~O.; Liu,~Z.~F. {Quasiparticle electronic structure of phthalocyanine:TMD interfaces from first-principles GW}. \emph{J. Chem. Phys.} \textbf{2021}, \emph{155}\relax
\mciteBstWouldAddEndPuncttrue
\mciteSetBstMidEndSepPunct{\mcitedefaultmidpunct}
{\mcitedefaultendpunct}{\mcitedefaultseppunct}\relax
\EndOfBibitem
\bibitem[Zhang \latin{et~al.}(2023)Zhang, Liu, Jiang, and Ma]{Zhang2023a}
Zhang,~M.; Liu,~Y.; Jiang,~Y.~N.; Ma,~Y. {Many-Body Green's Function Theory for Electronic Excitations in Complex Chemical Systems}. \emph{J. Phys. Chem. Lett.} \textbf{2023}, \emph{14}, 5267--5282\relax
\mciteBstWouldAddEndPuncttrue
\mciteSetBstMidEndSepPunct{\mcitedefaultmidpunct}
{\mcitedefaultendpunct}{\mcitedefaultseppunct}\relax
\EndOfBibitem
\bibitem[Umari \latin{et~al.}(2013)Umari, Giacomazzi, {De Angelis}, Pastore, and Baroni]{Umari2013}
Umari,~P.; Giacomazzi,~L.; {De Angelis},~F.; Pastore,~M.; Baroni,~S. {Energy-level alignment in organic dye-sensitized TiO2 from GW calculations}. \emph{J. Chem. Phys.} \textbf{2013}, \emph{139}, 014709\relax
\mciteBstWouldAddEndPuncttrue
\mciteSetBstMidEndSepPunct{\mcitedefaultmidpunct}
{\mcitedefaultendpunct}{\mcitedefaultseppunct}\relax
\EndOfBibitem
\bibitem[Marsili \latin{et~al.}(2017)Marsili, Mosconi, {De Angelis}, and Umari]{Marsili2017}
Marsili,~M.; Mosconi,~E.; {De Angelis},~F.; Umari,~P. {Large-scale GW-BSE calculations with N3 scaling: Excitonic effects in dye-sensitized solar cells}. \emph{Phys. Rev. B} \textbf{2017}, \emph{95}, 075415\relax
\mciteBstWouldAddEndPuncttrue
\mciteSetBstMidEndSepPunct{\mcitedefaultmidpunct}
{\mcitedefaultendpunct}{\mcitedefaultseppunct}\relax
\EndOfBibitem
\bibitem[Beli{\'{c}} \latin{et~al.}(2022)Beli{\'{c}}, F{\"{o}}rster, Menzel, Buda, and Visscher]{Belic2022}
Beli{\'{c}},~J.; F{\"{o}}rster,~A.; Menzel,~J.~P.; Buda,~F.; Visscher,~L. {Automated assessment of redox potentials for dyes in dye-sensitized photoelectrochemical cells}. \emph{Phys. Chem. Chem. Phys.} \textbf{2022}, \emph{24}, 197--210\relax
\mciteBstWouldAddEndPuncttrue
\mciteSetBstMidEndSepPunct{\mcitedefaultmidpunct}
{\mcitedefaultendpunct}{\mcitedefaultseppunct}\relax
\EndOfBibitem
\bibitem[Brooks \latin{et~al.}(2020)Brooks, Weng, Taylor, and Vlcek]{Brooks2020}
Brooks,~J.; Weng,~G.; Taylor,~S.; Vlcek,~V. {Stochastic many-body perturbation theory for Moir{\'{e}} states in twisted bilayer phosphorene}. \emph{J. Phys. Condens. Matter} \textbf{2020}, \emph{32}, 234001\relax
\mciteBstWouldAddEndPuncttrue
\mciteSetBstMidEndSepPunct{\mcitedefaultmidpunct}
{\mcitedefaultendpunct}{\mcitedefaultseppunct}\relax
\EndOfBibitem
\bibitem[Romanova and Vl{\v{c}}ek(2022)Romanova, and Vl{\v{c}}ek]{Romanova2022}
Romanova,~M.; Vl{\v{c}}ek,~V. {Stochastic many-body calculations of moir{\'{e}} states in twisted bilayer graphene at high pressures}. \emph{npj Comput. Mater.} \textbf{2022}, \emph{8}, 11\relax
\mciteBstWouldAddEndPuncttrue
\mciteSetBstMidEndSepPunct{\mcitedefaultmidpunct}
{\mcitedefaultendpunct}{\mcitedefaultseppunct}\relax
\EndOfBibitem
\bibitem[Graml \latin{et~al.}(2024)Graml, Zollner, Hernang, Junior, and Wilhelm]{Graml2024}
Graml,~M.; Zollner,~K.; Hernang,~D.; Junior,~P. E.~F.; Wilhelm,~J. {Low-Scaling GW Algorithm Applied to Twisted Transition-Metal Dichalcogenide Heterobilayers}. \emph{J. Chem. Theory Comput.} \textbf{2024}, \emph{20}, 2202--2208\relax
\mciteBstWouldAddEndPuncttrue
\mciteSetBstMidEndSepPunct{\mcitedefaultmidpunct}
{\mcitedefaultendpunct}{\mcitedefaultseppunct}\relax
\EndOfBibitem
\bibitem[Kaasbjerg and Thygesen(2010)Kaasbjerg, and Thygesen]{Kaasbjerg2010}
Kaasbjerg,~K.; Thygesen,~K.~S. {Benchmarking GW against exact diagonalization for semiempirical models}. \emph{Phys. Rev. B - Condens. Matter Mater. Phys.} \textbf{2010}, \emph{81}, 085102\relax
\mciteBstWouldAddEndPuncttrue
\mciteSetBstMidEndSepPunct{\mcitedefaultmidpunct}
{\mcitedefaultendpunct}{\mcitedefaultseppunct}\relax
\EndOfBibitem
\bibitem[van Loon \latin{et~al.}(2021)van Loon, R{\"{o}}sner, Katsnelson, and Wehling]{VanLoon2021}
van Loon,~E.~G.; R{\"{o}}sner,~M.; Katsnelson,~M.~I.; Wehling,~T.~O. {Random phase approximation for gapped systems: Role of vertex corrections and applicability of the constrained random phase approximation}. \emph{Phys. Rev. B} \textbf{2021}, \emph{104}, 045134\relax
\mciteBstWouldAddEndPuncttrue
\mciteSetBstMidEndSepPunct{\mcitedefaultmidpunct}
{\mcitedefaultendpunct}{\mcitedefaultseppunct}\relax
\EndOfBibitem
\bibitem[Bruneval and Marques(2013)Bruneval, and Marques]{Bruneval2013}
Bruneval,~F.; Marques,~M. {Benchmarking the starting points of the GW approximation for molecules}. \emph{J. Chem. Theory Comput.} \textbf{2013}, \emph{9}, 324--329\relax
\mciteBstWouldAddEndPuncttrue
\mciteSetBstMidEndSepPunct{\mcitedefaultmidpunct}
{\mcitedefaultendpunct}{\mcitedefaultseppunct}\relax
\EndOfBibitem
\bibitem[Vacondio \latin{et~al.}(2022)Vacondio, Varsano, Ruini, and Ferretti]{Vacondio2022}
Vacondio,~S.; Varsano,~D.; Ruini,~A.; Ferretti,~A. {Numerically Precise Benchmark of Many-Body Self-Energies on Spherical Atoms}. \emph{J. Chem. Theory Comput.} \textbf{2022}, \emph{18}, 3703--3717\relax
\mciteBstWouldAddEndPuncttrue
\mciteSetBstMidEndSepPunct{\mcitedefaultmidpunct}
{\mcitedefaultendpunct}{\mcitedefaultseppunct}\relax
\EndOfBibitem
\bibitem[Marie and Loos(2024)Marie, and Loos]{Marie2024}
Marie,~A.; Loos,~P.~F. {Reference Energies for Valence Ionizations and Satellite Transitions}. \emph{J. Chem. Theory Comput.} \textbf{2024}, \emph{20}, 4751--4777\relax
\mciteBstWouldAddEndPuncttrue
\mciteSetBstMidEndSepPunct{\mcitedefaultmidpunct}
{\mcitedefaultendpunct}{\mcitedefaultseppunct}\relax
\EndOfBibitem
\bibitem[Mahan and Sernelius(1989)Mahan, and Sernelius]{Mahan1989}
Mahan,~G.~D.; Sernelius,~B.~E. {Electron-Electron Interactions and the Bandwidth of Metals}. \emph{Phys. Rev. Lett.} \textbf{1989}, \emph{62}, 2718--2720\relax
\mciteBstWouldAddEndPuncttrue
\mciteSetBstMidEndSepPunct{\mcitedefaultmidpunct}
{\mcitedefaultendpunct}{\mcitedefaultseppunct}\relax
\EndOfBibitem
\bibitem[Shirley and Martin(1993)Shirley, and Martin]{Shirley1993}
Shirley,~E.~L.; Martin,~R.~M. {GW quasiparticle calculations in atoms}. \emph{Phys. Rev. B} \textbf{1993}, \emph{47}, 15404--15412\relax
\mciteBstWouldAddEndPuncttrue
\mciteSetBstMidEndSepPunct{\mcitedefaultmidpunct}
{\mcitedefaultendpunct}{\mcitedefaultseppunct}\relax
\EndOfBibitem
\bibitem[Bobbert and {Van Haeringen}(1994)Bobbert, and {Van Haeringen}]{Bobbert1994}
Bobbert,~P.~A.; {Van Haeringen},~W. {Lowest-order vertex-correction contribution to the direct gap of silicon}. \emph{Phys. Rev. B} \textbf{1994}, \emph{49}, 10326--10331\relax
\mciteBstWouldAddEndPuncttrue
\mciteSetBstMidEndSepPunct{\mcitedefaultmidpunct}
{\mcitedefaultendpunct}{\mcitedefaultseppunct}\relax
\EndOfBibitem
\bibitem[{Del Sole} \latin{et~al.}(1994){Del Sole}, Reining, and Godby]{delSole1994}
{Del Sole},~R.; Reining,~L.; Godby,~R. {GW$\Gamma$ approximation for electron self-energies in semiconductors and insulators}. \emph{Phys. Rev. B} \textbf{1994}, \emph{49}, 8024--8028\relax
\mciteBstWouldAddEndPuncttrue
\mciteSetBstMidEndSepPunct{\mcitedefaultmidpunct}
{\mcitedefaultendpunct}{\mcitedefaultseppunct}\relax
\EndOfBibitem
\bibitem[de~Groot \latin{et~al.}(1996)de~Groot, Ummels, Bobbert, and Haeringen]{DeGroot1996}
de~Groot,~H.; Ummels,~R.; Bobbert,~P.; Haeringen,~W.~V. {Lowest-order corrections to the RPA polarizability and GW self-energy of a semiconducting wire}. \emph{Phys. Rev. B - Condens. Matter Mater. Phys.} \textbf{1996}, \emph{54}, 2374--2380\relax
\mciteBstWouldAddEndPuncttrue
\mciteSetBstMidEndSepPunct{\mcitedefaultmidpunct}
{\mcitedefaultendpunct}{\mcitedefaultseppunct}\relax
\EndOfBibitem
\bibitem[Shirley(1996)]{Shirley1996}
Shirley,~E.~L. {Self-consistent GW and higher-order calculations of electron states in metals}. \emph{Phys. Rev. B - Condens. Matter Mater. Phys.} \textbf{1996}, \emph{54}, 7758--7764\relax
\mciteBstWouldAddEndPuncttrue
\mciteSetBstMidEndSepPunct{\mcitedefaultmidpunct}
{\mcitedefaultendpunct}{\mcitedefaultseppunct}\relax
\EndOfBibitem
\bibitem[Bechstedt \latin{et~al.}(1997)Bechstedt, Tenelsen, Adolph, and {Del Sole}]{Bechstedt1997}
Bechstedt,~F.; Tenelsen,~K.; Adolph,~B.; {Del Sole},~R. {Compensation of dynamical quasiparticle and vertex corrections in optical spectra}. \emph{Phys. Rev. Lett.} \textbf{1997}, \emph{78}, 1528--1531\relax
\mciteBstWouldAddEndPuncttrue
\mciteSetBstMidEndSepPunct{\mcitedefaultmidpunct}
{\mcitedefaultendpunct}{\mcitedefaultseppunct}\relax
\EndOfBibitem
\bibitem[Schindlmayr and Godby(1998)Schindlmayr, and Godby]{Schindlmayr1998}
Schindlmayr,~A.; Godby,~R.~W. {Systematic vertex corrections through iterative solution of Hedin's equations beyond the GW approximation}. \emph{Phys. Rev. Lett.} \textbf{1998}, \emph{80}, 1702--1705\relax
\mciteBstWouldAddEndPuncttrue
\mciteSetBstMidEndSepPunct{\mcitedefaultmidpunct}
{\mcitedefaultendpunct}{\mcitedefaultseppunct}\relax
\EndOfBibitem
\bibitem[Albrecht \latin{et~al.}(1998)Albrecht, Reining, {Del Sole}, and Onida]{Albrecht1998}
Albrecht,~S.; Reining,~L.; {Del Sole},~R.; Onida,~G. {Ab initio calculation of excitonic effects in the optical spectra of semiconductors}. \emph{Phys. Rev. Lett.} \textbf{1998}, \emph{80}, 4510\relax
\mciteBstWouldAddEndPuncttrue
\mciteSetBstMidEndSepPunct{\mcitedefaultmidpunct}
{\mcitedefaultendpunct}{\mcitedefaultseppunct}\relax
\EndOfBibitem
\bibitem[Ummels \latin{et~al.}(1998)Ummels, Bobbert, and van Haeringen]{Ummels1998}
Ummels,~R.; Bobbert,~P.~A.; van Haeringen,~W. {First-order corrections to random-phase approximation calculations in silicon and diamond}. \emph{Phys. Rev. B - Condens. Matter Mater. Phys.} \textbf{1998}, \emph{57}, 11962--11973\relax
\mciteBstWouldAddEndPuncttrue
\mciteSetBstMidEndSepPunct{\mcitedefaultmidpunct}
{\mcitedefaultendpunct}{\mcitedefaultseppunct}\relax
\EndOfBibitem
\bibitem[Schindlmayr \latin{et~al.}(1998)Schindlmayr, Pollehn, and Godby]{Schindlmayr1998a}
Schindlmayr,~A.; Pollehn,~T.~J.; Godby,~R. {Spectra and total energies from self-consistent many-body perturbation theory}. \emph{Phys. Rev. B - Condens. Matter Mater. Phys.} \textbf{1998}, \emph{58}, 12684--12690\relax
\mciteBstWouldAddEndPuncttrue
\mciteSetBstMidEndSepPunct{\mcitedefaultmidpunct}
{\mcitedefaultendpunct}{\mcitedefaultseppunct}\relax
\EndOfBibitem
\bibitem[Takada(2001)]{Takada2001}
Takada,~Y. {Inclusion of vertex corrections in the self-consistent calculation of quasiparticles in metals}. \emph{Phys. Rev. Lett.} \textbf{2001}, \emph{87}, 226402--226402--4\relax
\mciteBstWouldAddEndPuncttrue
\mciteSetBstMidEndSepPunct{\mcitedefaultmidpunct}
{\mcitedefaultendpunct}{\mcitedefaultseppunct}\relax
\EndOfBibitem
\bibitem[Bruneval \latin{et~al.}(2005)Bruneval, Sottile, Olevano, {Del Sole}, and Reining]{Bruneval2005}
Bruneval,~F.; Sottile,~F.; Olevano,~V.; {Del Sole},~R.; Reining,~L. {Many-body perturbation theory using the density-functional concept: Beyond the GW approximation}. \emph{Phys. Rev. Lett.} \textbf{2005}, \emph{94}, 186402\relax
\mciteBstWouldAddEndPuncttrue
\mciteSetBstMidEndSepPunct{\mcitedefaultmidpunct}
{\mcitedefaultendpunct}{\mcitedefaultseppunct}\relax
\EndOfBibitem
\bibitem[Morris \latin{et~al.}(2007)Morris, Stankovski, Delaney, Rinke, Garc{\'{i}}a-Gonz{\'{a}}lez, and Godby]{Morris2007}
Morris,~A.~J.; Stankovski,~M.; Delaney,~K.~T.; Rinke,~P.; Garc{\'{i}}a-Gonz{\'{a}}lez,~P.; Godby,~R.~W. {Vertex corrections in localized and extended systems}. \emph{Phys. Rev. B - Condens. Matter Mater. Phys.} \textbf{2007}, \emph{76}, 155106\relax
\mciteBstWouldAddEndPuncttrue
\mciteSetBstMidEndSepPunct{\mcitedefaultmidpunct}
{\mcitedefaultendpunct}{\mcitedefaultseppunct}\relax
\EndOfBibitem
\bibitem[Shishkin \latin{et~al.}(2007)Shishkin, Marsman, and Kresse]{Shishkin2007}
Shishkin,~M.; Marsman,~M.; Kresse,~G. {Accurate quasiparticle spectra from self-consistent GW calculations with vertex corrections}. \emph{Phys. Rev. Lett.} \textbf{2007}, \emph{99}, 246403\relax
\mciteBstWouldAddEndPuncttrue
\mciteSetBstMidEndSepPunct{\mcitedefaultmidpunct}
{\mcitedefaultendpunct}{\mcitedefaultseppunct}\relax
\EndOfBibitem
\bibitem[Gr{\"{u}}neis \latin{et~al.}(2014)Gr{\"{u}}neis, Kresse, Hinuma, and Oba]{Gruneis2014}
Gr{\"{u}}neis,~A.; Kresse,~G.; Hinuma,~Y.; Oba,~F. {Ionization potentials of solids: The importance of vertex corrections}. \emph{Phys. Rev. Lett.} \textbf{2014}, \emph{112}, 096401\relax
\mciteBstWouldAddEndPuncttrue
\mciteSetBstMidEndSepPunct{\mcitedefaultmidpunct}
{\mcitedefaultendpunct}{\mcitedefaultseppunct}\relax
\EndOfBibitem
\bibitem[Stefanucci \latin{et~al.}(2014)Stefanucci, Pavlyukh, Uimonen, and van Leeuwen]{Stefanucci2014}
Stefanucci,~G.; Pavlyukh,~Y.; Uimonen,~A.~M.; van Leeuwen,~R. {Diagrammatic expansion for positive spectral functions beyond GW: Application to vertex corrections in the electron gas}. \emph{Phys. Rev. B} \textbf{2014}, \emph{90}, 115134\relax
\mciteBstWouldAddEndPuncttrue
\mciteSetBstMidEndSepPunct{\mcitedefaultmidpunct}
{\mcitedefaultendpunct}{\mcitedefaultseppunct}\relax
\EndOfBibitem
\bibitem[Ren \latin{et~al.}(2015)Ren, Marom, Caruso, Scheffler, and Rinke]{Ren2015}
Ren,~X.; Marom,~N.; Caruso,~F.; Scheffler,~M.; Rinke,~P. {Beyond the GW approximation: A second-order screened exchange correction}. \emph{Phys. Rev. B - Condens. Matter Mater. Phys.} \textbf{2015}, \emph{92}, 081104(R)\relax
\mciteBstWouldAddEndPuncttrue
\mciteSetBstMidEndSepPunct{\mcitedefaultmidpunct}
{\mcitedefaultendpunct}{\mcitedefaultseppunct}\relax
\EndOfBibitem
\bibitem[Chen and Pasquarello(2015)Chen, and Pasquarello]{Chen2015}
Chen,~W.; Pasquarello,~A. {Accurate band gaps of extended systems via efficient vertex corrections in GW}. \emph{Phys. Rev. B - Condens. Matter Mater. Phys.} \textbf{2015}, \emph{92}, 041115(R)\relax
\mciteBstWouldAddEndPuncttrue
\mciteSetBstMidEndSepPunct{\mcitedefaultmidpunct}
{\mcitedefaultendpunct}{\mcitedefaultseppunct}\relax
\EndOfBibitem
\bibitem[Kutepov(2016)]{Kutepov2016}
Kutepov,~A.~L. {Electronic structure of Na, K, Si, and LiF from self-consistent solution of Hedin's equations including vertex corrections}. \emph{Phys. Rev. B} \textbf{2016}, \emph{94}, 155101\relax
\mciteBstWouldAddEndPuncttrue
\mciteSetBstMidEndSepPunct{\mcitedefaultmidpunct}
{\mcitedefaultendpunct}{\mcitedefaultseppunct}\relax
\EndOfBibitem
\bibitem[Pavlyukh \latin{et~al.}(2016)Pavlyukh, Uimonen, Stefanucci, and van Leeuwen]{Pavlyukh2016}
Pavlyukh,~Y.; Uimonen,~A.~M.; Stefanucci,~G.; van Leeuwen,~R. {Vertex corrections for positive-definite spectral functions of simple metals}. \emph{Phys. Rev. Lett.} \textbf{2016}, \emph{117}, 206402\relax
\mciteBstWouldAddEndPuncttrue
\mciteSetBstMidEndSepPunct{\mcitedefaultmidpunct}
{\mcitedefaultendpunct}{\mcitedefaultseppunct}\relax
\EndOfBibitem
\bibitem[Hung \latin{et~al.}(2016)Hung, da~Jornada, Souto-Casares, Chelikowsky, Louie, and {\"{O}}ǧ{\"{u}}t]{Hung2016}
Hung,~L.; da~Jornada,~F.~H.; Souto-Casares,~J.; Chelikowsky,~J.~R.; Louie,~S.~G.; {\"{O}}ǧ{\"{u}}t,~S. {Excitation spectra of aromatic molecules within a real-space GW -BSE formalism: Role of self-consistency and vertex corrections}. \emph{Phys. Rev. B} \textbf{2016}, \emph{94}, 085125\relax
\mciteBstWouldAddEndPuncttrue
\mciteSetBstMidEndSepPunct{\mcitedefaultmidpunct}
{\mcitedefaultendpunct}{\mcitedefaultseppunct}\relax
\EndOfBibitem
\bibitem[Kuwahara \latin{et~al.}(2016)Kuwahara, Noguchi, and Ohno]{Kuwahara2016}
Kuwahara,~R.; Noguchi,~Y.; Ohno,~K. {GW $\Gamma$ + Bethe-Salpeter equation approach for photoabsorption spectra: Importance of self-consistent GW $\Gamma$ calculations in small atomic systems}. \emph{Phys. Rev. B} \textbf{2016}, \emph{94}, 121116(R)\relax
\mciteBstWouldAddEndPuncttrue
\mciteSetBstMidEndSepPunct{\mcitedefaultmidpunct}
{\mcitedefaultendpunct}{\mcitedefaultseppunct}\relax
\EndOfBibitem
\bibitem[Kutepov(2017)]{Kutepov2017a}
Kutepov,~A.~L. {Self-consistent solution of Hedin's equations: Semiconductors and insulators}. \emph{Phys. Rev. B} \textbf{2017}, \emph{95}, 195120\relax
\mciteBstWouldAddEndPuncttrue
\mciteSetBstMidEndSepPunct{\mcitedefaultmidpunct}
{\mcitedefaultendpunct}{\mcitedefaultseppunct}\relax
\EndOfBibitem
\bibitem[Schmidt \latin{et~al.}(2017)Schmidt, Patrick, and Thygesen]{Schmidt2017}
Schmidt,~P.~S.; Patrick,~C.~E.; Thygesen,~K.~S. {Simple vertex correction improves GW band energies of bulk and two-dimensional crystals}. \emph{Phys. Rev. B} \textbf{2017}, \emph{96}, 205206\relax
\mciteBstWouldAddEndPuncttrue
\mciteSetBstMidEndSepPunct{\mcitedefaultmidpunct}
{\mcitedefaultendpunct}{\mcitedefaultseppunct}\relax
\EndOfBibitem
\bibitem[Maggio and Kresse(2017)Maggio, and Kresse]{Maggio2017}
Maggio,~E.; Kresse,~G. {GW Vertex Corrected Calculations for Molecular Systems}. \emph{J. Chem. Theory Comput.} \textbf{2017}, \emph{13}, 4765--4778\relax
\mciteBstWouldAddEndPuncttrue
\mciteSetBstMidEndSepPunct{\mcitedefaultmidpunct}
{\mcitedefaultendpunct}{\mcitedefaultseppunct}\relax
\EndOfBibitem
\bibitem[Maggio and Kresse(2018)Maggio, and Kresse]{Maggio2018}
Maggio,~E.; Kresse,~G. {Correction to: GW Vertex corrected calculations for molecular systems}. \emph{J. Chem. Theory Comput.} \textbf{2018}, \emph{14}, 1821\relax
\mciteBstWouldAddEndPuncttrue
\mciteSetBstMidEndSepPunct{\mcitedefaultmidpunct}
{\mcitedefaultendpunct}{\mcitedefaultseppunct}\relax
\EndOfBibitem
\bibitem[Kutepov and Kotliar(2017)Kutepov, and Kotliar]{Kutepov2017}
Kutepov,~A.~L.; Kotliar,~G. {One-electron spectra and susceptibilities of the three-dimensional electron gas from self-consistent solutions of Hedin's equations}. \emph{Phys. Rev. B} \textbf{2017}, \emph{96}, 035108\relax
\mciteBstWouldAddEndPuncttrue
\mciteSetBstMidEndSepPunct{\mcitedefaultmidpunct}
{\mcitedefaultendpunct}{\mcitedefaultseppunct}\relax
\EndOfBibitem
\bibitem[Cunningham \latin{et~al.}(2018)Cunningham, Gr{\"{u}}ning, Azarhoosh, Pashov, and van Schilfgaarde]{Cunningham2018}
Cunningham,~B.; Gr{\"{u}}ning,~M.; Azarhoosh,~P.; Pashov,~D.; van Schilfgaarde,~M. {Effect of ladder diagrams on optical absorption spectra in a quasiparticle self-consistent GW framework}. \emph{Phys. Rev. Mater.} \textbf{2018}, \emph{2}, 034603\relax
\mciteBstWouldAddEndPuncttrue
\mciteSetBstMidEndSepPunct{\mcitedefaultmidpunct}
{\mcitedefaultendpunct}{\mcitedefaultseppunct}\relax
\EndOfBibitem
\bibitem[Olsen \latin{et~al.}(2019)Olsen, Patrick, Bates, Ruzsinszky, and Thygesen]{Olsen2019}
Olsen,~T.; Patrick,~C.~E.; Bates,~J.~E.; Ruzsinszky,~A.; Thygesen,~K.~S. {Beyond the RPA and GW methods with adiabatic xc-kernels for accurate ground state and quasiparticle energies}. \emph{Nat. Comput. Mater.} \textbf{2019}, \emph{5}, 106\relax
\mciteBstWouldAddEndPuncttrue
\mciteSetBstMidEndSepPunct{\mcitedefaultmidpunct}
{\mcitedefaultendpunct}{\mcitedefaultseppunct}\relax
\EndOfBibitem
\bibitem[Vl{\v{c}}ek(2019)]{Vlcek2019}
Vl{\v{c}}ek,~V. {Stochastic Vertex Corrections: Linear Scaling Methods for Accurate Quasiparticle Energies}. \emph{J. Chem. Theory Comput.} \textbf{2019}, \emph{15}, 6254--6266\relax
\mciteBstWouldAddEndPuncttrue
\mciteSetBstMidEndSepPunct{\mcitedefaultmidpunct}
{\mcitedefaultendpunct}{\mcitedefaultseppunct}\relax
\EndOfBibitem
\bibitem[Lewis and Berkelbach(2019)Lewis, and Berkelbach]{Lewis2019}
Lewis,~A.~M.; Berkelbach,~T.~C. {Vertex Corrections to the Polarizability Do Not Improve the GW Approximation for the Ionization Potential of Molecules}. \emph{J. Chem. Theory Comput.} \textbf{2019}, \emph{15}, 2925--2932\relax
\mciteBstWouldAddEndPuncttrue
\mciteSetBstMidEndSepPunct{\mcitedefaultmidpunct}
{\mcitedefaultendpunct}{\mcitedefaultseppunct}\relax
\EndOfBibitem
\bibitem[Pavlyukh \latin{et~al.}(2020)Pavlyukh, Stefanucci, and van Leeuwen]{Pavlyukh2020}
Pavlyukh,~Y.; Stefanucci,~G.; van Leeuwen,~R. {Dynamically screened vertex correction to GW}. \emph{Phys. Rev. B} \textbf{2020}, \emph{102}, 045121\relax
\mciteBstWouldAddEndPuncttrue
\mciteSetBstMidEndSepPunct{\mcitedefaultmidpunct}
{\mcitedefaultendpunct}{\mcitedefaultseppunct}\relax
\EndOfBibitem
\bibitem[Wang \latin{et~al.}(2021)Wang, Rinke, and Ren]{Wang2021}
Wang,~Y.; Rinke,~P.; Ren,~X. {Assessing the G0W0$\Gamma$0(1) Approach: Beyond G0W0 with Hedin's Full Second-Order Self-Energy Contribution}. \emph{J. Chem. Theory Comput.} \textbf{2021}, \emph{17}, 5140--5154\relax
\mciteBstWouldAddEndPuncttrue
\mciteSetBstMidEndSepPunct{\mcitedefaultmidpunct}
{\mcitedefaultendpunct}{\mcitedefaultseppunct}\relax
\EndOfBibitem
\bibitem[Mejuto-Zaera \latin{et~al.}(2021)Mejuto-Zaera, Weng, Romanova, Cotton, Whaley, Tubman, and Vl{\v{c}}ek]{Mejuto-Zaera2021}
Mejuto-Zaera,~C.; Weng,~G.; Romanova,~M.; Cotton,~S.~J.; Whaley,~K.~B.; Tubman,~N.~M.; Vl{\v{c}}ek,~V. {Are multi-quasiparticle interactions important in molecular ionization?} \emph{J. Chem. Phys.} \textbf{2021}, \emph{154}, 121101\relax
\mciteBstWouldAddEndPuncttrue
\mciteSetBstMidEndSepPunct{\mcitedefaultmidpunct}
{\mcitedefaultendpunct}{\mcitedefaultseppunct}\relax
\EndOfBibitem
\bibitem[Tal \latin{et~al.}(2021)Tal, Chen, and Pasquarello]{Tal2021}
Tal,~A.; Chen,~W.; Pasquarello,~A. {Vertex function compliant with the Ward identity for quasiparticle self-consistent calculations beyond GW}. \emph{Phys. Rev. B} \textbf{2021}, \emph{103}, 161104\relax
\mciteBstWouldAddEndPuncttrue
\mciteSetBstMidEndSepPunct{\mcitedefaultmidpunct}
{\mcitedefaultendpunct}{\mcitedefaultseppunct}\relax
\EndOfBibitem
\bibitem[Joost \latin{et~al.}(2022)Joost, Schl{\"{u}}nzen, Ohldag, Bonitz, Lackner, and Březinov{\'{a}}]{Joost2022a}
Joost,~J.~P.; Schl{\"{u}}nzen,~N.; Ohldag,~H.; Bonitz,~M.; Lackner,~F.; Březinov{\'{a}},~I. {Dynamically screened ladder approximation: Simultaneous treatment of strong electronic correlations and dynamical screening out of equilibrium}. \emph{Phys. Rev. B} \textbf{2022}, \emph{105}, 165155\relax
\mciteBstWouldAddEndPuncttrue
\mciteSetBstMidEndSepPunct{\mcitedefaultmidpunct}
{\mcitedefaultendpunct}{\mcitedefaultseppunct}\relax
\EndOfBibitem
\bibitem[Mejuto-Zaera and Vl{\v{c}}ek(2022)Mejuto-Zaera, and Vl{\v{c}}ek]{Mejuto-Zaera2022a}
Mejuto-Zaera,~C.; Vl{\v{c}}ek,~V. {Self-consistency in GW $\Gamma$ formalism leading to quasiparticle-quasiparticle couplings}. \emph{Phys. Rev. B} \textbf{2022}, \emph{106}, 165129\relax
\mciteBstWouldAddEndPuncttrue
\mciteSetBstMidEndSepPunct{\mcitedefaultmidpunct}
{\mcitedefaultendpunct}{\mcitedefaultseppunct}\relax
\EndOfBibitem
\bibitem[F{\"{o}}rster and Visscher(2022)F{\"{o}}rster, and Visscher]{Forster2022}
F{\"{o}}rster,~A.; Visscher,~L. {Exploring the statically screened G3W2 correction to the GW self-energy : Charged excitations and total energies of finite systems}. \emph{Phys. Rev. B} \textbf{2022}, \emph{105}, 125121\relax
\mciteBstWouldAddEndPuncttrue
\mciteSetBstMidEndSepPunct{\mcitedefaultmidpunct}
{\mcitedefaultendpunct}{\mcitedefaultseppunct}\relax
\EndOfBibitem
\bibitem[Rohlfing(2023)]{Rohlfing2023}
Rohlfing,~M. {Approximate spatiotemporal structure of the vertex function $\Gamma$⁡( 1, 2;3) in many-body perturbation theory}. \emph{Phys. Rev. B} \textbf{2023}, \emph{108}, 195207\relax
\mciteBstWouldAddEndPuncttrue
\mciteSetBstMidEndSepPunct{\mcitedefaultmidpunct}
{\mcitedefaultendpunct}{\mcitedefaultseppunct}\relax
\EndOfBibitem
\bibitem[Lorin \latin{et~al.}(2023)Lorin, Bischoff, Tal, and Pasquarello]{Lorin2023}
Lorin,~A.; Bischoff,~T.; Tal,~A.; Pasquarello,~A. {Band alignments through quasiparticle self-consistent GW with efficient vertex corrections}. \emph{Phys. Rev. B} \textbf{2023}, \emph{108}, 245303\relax
\mciteBstWouldAddEndPuncttrue
\mciteSetBstMidEndSepPunct{\mcitedefaultmidpunct}
{\mcitedefaultendpunct}{\mcitedefaultseppunct}\relax
\EndOfBibitem
\bibitem[Vacondio \latin{et~al.}(2024)Vacondio, Varsano, Ruini, and Ferretti]{Vacondio2024}
Vacondio,~S.; Varsano,~D.; Ruini,~A.; Ferretti,~A. {Going Beyond the GW Approximation Using the Time-Dependent Hartree−Fock Vertex}. \emph{J. Chem. Theory Comput.} \textbf{2024}, \emph{20}, 4718--4737\relax
\mciteBstWouldAddEndPuncttrue
\mciteSetBstMidEndSepPunct{\mcitedefaultmidpunct}
{\mcitedefaultendpunct}{\mcitedefaultseppunct}\relax
\EndOfBibitem
\bibitem[Tal \latin{et~al.}(2024)Tal, Bischoff, and Pasquarello]{Tal2024}
Tal,~A.; Bischoff,~T.; Pasquarello,~A. {Absolute energy levels of liquid water from many-body perturbation theory with effective vertex corrections}. \emph{Proc. Natl. Acad. Sci.} \textbf{2024}, \emph{121}, e2311472121\relax
\mciteBstWouldAddEndPuncttrue
\mciteSetBstMidEndSepPunct{\mcitedefaultmidpunct}
{\mcitedefaultendpunct}{\mcitedefaultseppunct}\relax
\EndOfBibitem
\bibitem[Abdallah and Pasquarello(2024)Abdallah, and Pasquarello]{Abdallah2024}
Abdallah,~M.~S.; Pasquarello,~A. {Quasiparticle self-consistent GW with effective vertex corrections in the polarizability and the self-energy applied to MnO, FeO, CoO, and NiO}. \emph{Phys. Rev. B} \textbf{2024}, \emph{110}, 155105\relax
\mciteBstWouldAddEndPuncttrue
\mciteSetBstMidEndSepPunct{\mcitedefaultmidpunct}
{\mcitedefaultendpunct}{\mcitedefaultseppunct}\relax
\EndOfBibitem
\bibitem[Wen \latin{et~al.}(2024)Wen, Abraham, Harsha, Shee, Whaley, and Zgid]{Wen2024}
Wen,~M.; Abraham,~V.; Harsha,~G.; Shee,~A.; Whaley,~K.~B.; Zgid,~D. {Comparing Self-Consistent GW and Vertex-Corrected G0W0 (G0W0$\Gamma$) Accuracy for Molecular Ionization Potentials}. \emph{J. Chem. Theory Comput.} \textbf{2024}, \emph{20}, 3109--3120\relax
\mciteBstWouldAddEndPuncttrue
\mciteSetBstMidEndSepPunct{\mcitedefaultmidpunct}
{\mcitedefaultendpunct}{\mcitedefaultseppunct}\relax
\EndOfBibitem
\bibitem[Patterson(2024)]{Patterson2024}
Patterson,~C.~H. {Molecular Ionization Energies from GW and Hartree−Fock Theory: Polarizability, Screening, and Self-Energy Vertex Corrections}. \emph{J. Chem. Theory Comput.} \textbf{2024}, \emph{20}, 7479--7493\relax
\mciteBstWouldAddEndPuncttrue
\mciteSetBstMidEndSepPunct{\mcitedefaultmidpunct}
{\mcitedefaultendpunct}{\mcitedefaultseppunct}\relax
\EndOfBibitem
\bibitem[Bruneval and F{\"{o}}rster(2024)Bruneval, and F{\"{o}}rster]{Bruneval2024}
Bruneval,~F.; F{\"{o}}rster,~A. {Fully dynamic G3W2 self-energy for finite systems: Formulas and benchmark}. \emph{J. Chem. Theory Comput.} \textbf{2024}, \emph{20}, 3218−3230\relax
\mciteBstWouldAddEndPuncttrue
\mciteSetBstMidEndSepPunct{\mcitedefaultmidpunct}
{\mcitedefaultendpunct}{\mcitedefaultseppunct}\relax
\EndOfBibitem
\bibitem[Kutepov(2021)]{Kutepov2021c}
Kutepov,~A.~L. {Electronic structure of van der Waals ferromagnet CrI3 from self consistent vertex corrected GW approaches}. \emph{Phys. Rev. Mater.} \textbf{2021}, \emph{5}, 083805\relax
\mciteBstWouldAddEndPuncttrue
\mciteSetBstMidEndSepPunct{\mcitedefaultmidpunct}
{\mcitedefaultendpunct}{\mcitedefaultseppunct}\relax
\EndOfBibitem
\bibitem[Kutepov(2022)]{Kutepov2022}
Kutepov,~A.~L. {Full versus quasiparticle self-consistency in vertex-corrected GW approaches}. \emph{Phys. Rev. B} \textbf{2022}, \emph{105}, 045124\relax
\mciteBstWouldAddEndPuncttrue
\mciteSetBstMidEndSepPunct{\mcitedefaultmidpunct}
{\mcitedefaultendpunct}{\mcitedefaultseppunct}\relax
\EndOfBibitem
\bibitem[Knight \latin{et~al.}(2016)Knight, Wang, Gallandi, Dolgounitcheva, Ren, Ortiz, Rinke, K{\"{o}}rzd{\"{o}}rfer, and Marom]{Knight2016}
Knight,~J.~W.; Wang,~X.; Gallandi,~L.; Dolgounitcheva,~O.; Ren,~X.; Ortiz,~J.~V.; Rinke,~P.; K{\"{o}}rzd{\"{o}}rfer,~T.; Marom,~N. {Accurate Ionization Potentials and Electron Affinities of Acceptor Molecules III: A Benchmark of GW Methods}. \emph{J. Chem. Theory Comput.} \textbf{2016}, \emph{12}, 615--626\relax
\mciteBstWouldAddEndPuncttrue
\mciteSetBstMidEndSepPunct{\mcitedefaultmidpunct}
{\mcitedefaultendpunct}{\mcitedefaultseppunct}\relax
\EndOfBibitem
\bibitem[Bruneval \latin{et~al.}(2021)Bruneval, Dattani, and van Setten]{Bruneval2021a}
Bruneval,~F.; Dattani,~N.; van Setten,~M.~J. {The GW Miracle in Many-Body Perturbation Theory for the Ionization Potential of Molecules}. \emph{Front. Chem.} \textbf{2021}, \emph{9}, 749779\relax
\mciteBstWouldAddEndPuncttrue
\mciteSetBstMidEndSepPunct{\mcitedefaultmidpunct}
{\mcitedefaultendpunct}{\mcitedefaultseppunct}\relax
\EndOfBibitem
\bibitem[McKeon \latin{et~al.}(2022)McKeon, Hamed, Bruneval, and Neaton]{McKeon2022}
McKeon,~C.~A.; Hamed,~S.~M.; Bruneval,~F.; Neaton,~J.~B. {An optimally tuned range-separated hybrid starting point for ab initio GW plus Bethe-Salpeter equation calculations of molecules}. \emph{J. Chem. Phys.} \textbf{2022}, \emph{157}, 074103\relax
\mciteBstWouldAddEndPuncttrue
\mciteSetBstMidEndSepPunct{\mcitedefaultmidpunct}
{\mcitedefaultendpunct}{\mcitedefaultseppunct}\relax
\EndOfBibitem
\bibitem[Kotani \latin{et~al.}(2007)Kotani, van Schilfgaarde, and Faleev]{Kotani2007}
Kotani,~T.; van Schilfgaarde,~M.; Faleev,~S.~V. {Quasiparticle self-consistent GW method: A basis for the independent-particle approximation}. \emph{Phys. Rev. B} \textbf{2007}, \emph{76}, 165106\relax
\mciteBstWouldAddEndPuncttrue
\mciteSetBstMidEndSepPunct{\mcitedefaultmidpunct}
{\mcitedefaultendpunct}{\mcitedefaultseppunct}\relax
\EndOfBibitem
\bibitem[Romaniello \latin{et~al.}(2009)Romaniello, Guyot, and Reining]{romaniello_jchemphys2009}
Romaniello,~P.; Guyot,~S.; Reining,~L. {The self-energy beyond GW: Local and nonlocal vertex corrections}. \emph{J. Chem. Phys.} \textbf{2009}, \emph{131}, 154111\relax
\mciteBstWouldAddEndPuncttrue
\mciteSetBstMidEndSepPunct{\mcitedefaultmidpunct}
{\mcitedefaultendpunct}{\mcitedefaultseppunct}\relax
\EndOfBibitem
\bibitem[Chang and Jin(2012)Chang, and Jin]{Chang2012}
Chang,~Y.~W.; Jin,~B.~Y. {Self-interaction correction to GW approximation}. \emph{Phys. Scr.} \textbf{2012}, \emph{86}, 065301\relax
\mciteBstWouldAddEndPuncttrue
\mciteSetBstMidEndSepPunct{\mcitedefaultmidpunct}
{\mcitedefaultendpunct}{\mcitedefaultseppunct}\relax
\EndOfBibitem
\bibitem[Patterson(2024)]{Patterson2024a}
Patterson,~C.~H. {Erratum: Molecular Ionization Energies from GW and Hartree-Fock Theory: Polarizability, Screening and Self-Energy Vertex Corrections}. \emph{J. Chem. Theory Comput.} \textbf{2024}, \emph{20}, 9267\relax
\mciteBstWouldAddEndPuncttrue
\mciteSetBstMidEndSepPunct{\mcitedefaultmidpunct}
{\mcitedefaultendpunct}{\mcitedefaultseppunct}\relax
\EndOfBibitem
\bibitem[Cunningham \latin{et~al.}(2023)Cunningham, Gr{\"{u}}ning, Pashov, and {Van Schilfgaarde}]{Cunningham2023}
Cunningham,~B.; Gr{\"{u}}ning,~M.; Pashov,~D.; {Van Schilfgaarde},~M. {QSG Ŵ: Quasiparticle self-consistent GW with ladder diagrams in W}. \emph{Phys. Rev. B} \textbf{2023}, \emph{108}, 165104\relax
\mciteBstWouldAddEndPuncttrue
\mciteSetBstMidEndSepPunct{\mcitedefaultmidpunct}
{\mcitedefaultendpunct}{\mcitedefaultseppunct}\relax
\EndOfBibitem
\bibitem[van Schilfgaarde \latin{et~al.}(2006)van Schilfgaarde, Kotani, and Faleev]{VanSchilfgaarde2006}
van Schilfgaarde,~M.; Kotani,~T.; Faleev,~S. {Quasiparticle self-consistent GW theory}. \emph{Phys. Rev. Lett.} \textbf{2006}, \emph{96}, 226402\relax
\mciteBstWouldAddEndPuncttrue
\mciteSetBstMidEndSepPunct{\mcitedefaultmidpunct}
{\mcitedefaultendpunct}{\mcitedefaultseppunct}\relax
\EndOfBibitem
\bibitem[Faleev \latin{et~al.}(2004)Faleev, van Schilfgaarde, and Kotani]{Faleev2004}
Faleev,~S.~V.; van Schilfgaarde,~M.; Kotani,~T. {All-electron self-consistent GW approximation: Application to Si, MnO, and NiO}. \emph{Phys. Rev. Lett.} \textbf{2004}, \emph{93}, 126406\relax
\mciteBstWouldAddEndPuncttrue
\mciteSetBstMidEndSepPunct{\mcitedefaultmidpunct}
{\mcitedefaultendpunct}{\mcitedefaultseppunct}\relax
\EndOfBibitem
\bibitem[Bruneval and Gatti(2014)Bruneval, and Gatti]{bruneval_springer2014}
Bruneval,~F.; Gatti,~M. In \emph{First Principles Approaches to Spectroscopic Properties of Complex Materials}; Di~Valentin,~C., Botti,~S., Cococcioni,~M., Eds.; Topics in Current Chemistry; Springer Berlin Heidelberg, 2014; Vol. 347; pp 99--136\relax
\mciteBstWouldAddEndPuncttrue
\mciteSetBstMidEndSepPunct{\mcitedefaultmidpunct}
{\mcitedefaultendpunct}{\mcitedefaultseppunct}\relax
\EndOfBibitem
\bibitem[Bintrim and Berkelbach(2021)Bintrim, and Berkelbach]{Bintrim2021}
Bintrim,~S.~J.; Berkelbach,~T.~C. {Full-frequency GW without frequency}. \emph{J. Chem. Phys.} \textbf{2021}, \emph{154}, 041101\relax
\mciteBstWouldAddEndPuncttrue
\mciteSetBstMidEndSepPunct{\mcitedefaultmidpunct}
{\mcitedefaultendpunct}{\mcitedefaultseppunct}\relax
\EndOfBibitem
\bibitem[Strinati(1988)]{Strinati1988}
Strinati,~G. {Application of the Green's functions method to the study of the optical properties of semiconductors}. \emph{La Riv. Del Nuovo Cim. Ser. 3} \textbf{1988}, \emph{11}, 1--86\relax
\mciteBstWouldAddEndPuncttrue
\mciteSetBstMidEndSepPunct{\mcitedefaultmidpunct}
{\mcitedefaultendpunct}{\mcitedefaultseppunct}\relax
\EndOfBibitem
\bibitem[Romaniello \latin{et~al.}(2012)Romaniello, Bechstedt, and Reining]{Romaniello2012}
Romaniello,~P.; Bechstedt,~F.; Reining,~L. {Beyond the GW approximation: Combining correlation channels}. \emph{Phys. Rev. B} \textbf{2012}, \emph{85}, 155131\relax
\mciteBstWouldAddEndPuncttrue
\mciteSetBstMidEndSepPunct{\mcitedefaultmidpunct}
{\mcitedefaultendpunct}{\mcitedefaultseppunct}\relax
\EndOfBibitem
\bibitem[Orlando \latin{et~al.}(2023)Orlando, Romaniello, and Loos]{Orlando2023}
Orlando,~R.; Romaniello,~P.; Loos,~P.~F. {The three channels of many-body perturbation theory: GW, particle-particle, and electron-hole T-matrix self-energies}. \emph{J. Chem. Phys.} \textbf{2023}, \emph{159}, 184113\relax
\mciteBstWouldAddEndPuncttrue
\mciteSetBstMidEndSepPunct{\mcitedefaultmidpunct}
{\mcitedefaultendpunct}{\mcitedefaultseppunct}\relax
\EndOfBibitem
\bibitem[Starke and Kresse(2012)Starke, and Kresse]{Starke2012}
Starke,~R.; Kresse,~G. {Self-consistent Green function equations and the hierarchy of approximations for the four-point propagator}. \emph{Phys. Rev. B} \textbf{2012}, \emph{85}, 075119\relax
\mciteBstWouldAddEndPuncttrue
\mciteSetBstMidEndSepPunct{\mcitedefaultmidpunct}
{\mcitedefaultendpunct}{\mcitedefaultseppunct}\relax
\EndOfBibitem
\bibitem[D{\'{i}}az–Tinoco \latin{et~al.}(2019)D{\'{i}}az–Tinoco, Corzo, Paw{\l}owski, and Ortiz]{DiazTinoco2019}
D{\'{i}}az–Tinoco,~M.; Corzo,~H.~H.; Paw{\l}owski,~F.; Ortiz,~J.~V. {Do Dyson Orbitals resemble canonical Hartree–Fock orbitals?} \emph{Mol. Phys.} \textbf{2019}, \emph{117}, 2275--2283\relax
\mciteBstWouldAddEndPuncttrue
\mciteSetBstMidEndSepPunct{\mcitedefaultmidpunct}
{\mcitedefaultendpunct}{\mcitedefaultseppunct}\relax
\EndOfBibitem
\bibitem[Romaniello \latin{et~al.}(2009)Romaniello, Sangalli, Berger, Sottile, Molinari, Reining, and Onida]{Romaniello2009b}
Romaniello,~P.; Sangalli,~D.; Berger,~J.~A.; Sottile,~F.; Molinari,~L.~G.; Reining,~L.; Onida,~G. {Double excitations in finite systems}. \emph{J. Chem. Phys.} \textbf{2009}, \emph{130}, 044108\relax
\mciteBstWouldAddEndPuncttrue
\mciteSetBstMidEndSepPunct{\mcitedefaultmidpunct}
{\mcitedefaultendpunct}{\mcitedefaultseppunct}\relax
\EndOfBibitem
\bibitem[Rohlfing and Louie(2000)Rohlfing, and Louie]{Rohlfing2000}
Rohlfing,~M.; Louie,~S.~G. {Electron-hole excitations and optical spectra from first principles}. \emph{Phys. Rev. B} \textbf{2000}, \emph{62}, 4927--4944\relax
\mciteBstWouldAddEndPuncttrue
\mciteSetBstMidEndSepPunct{\mcitedefaultmidpunct}
{\mcitedefaultendpunct}{\mcitedefaultseppunct}\relax
\EndOfBibitem
\bibitem[Ullrich(2012)]{ullrich_book}
Ullrich,~C. \emph{Time-Dependent Density-Functional Theory}; Oxford University Press: New York, 2012\relax
\mciteBstWouldAddEndPuncttrue
\mciteSetBstMidEndSepPunct{\mcitedefaultmidpunct}
{\mcitedefaultendpunct}{\mcitedefaultseppunct}\relax
\EndOfBibitem
\bibitem[Blase \latin{et~al.}(2020)Blase, Duchemin, Jacquemin, and Loos]{blase_jpcl2020}
Blase,~X.; Duchemin,~I.; Jacquemin,~D.; Loos,~P.-F. The Bethe–Salpeter Equation Formalism: From Physics to Chemistry. \emph{The Journal of Physical Chemistry Letters} \textbf{2020}, \emph{11}, 7371--7382, PMID: 32787315\relax
\mciteBstWouldAddEndPuncttrue
\mciteSetBstMidEndSepPunct{\mcitedefaultmidpunct}
{\mcitedefaultendpunct}{\mcitedefaultseppunct}\relax
\EndOfBibitem
\bibitem[Irmler \latin{et~al.}(2019)Irmler, Gallo, Hummel, and Gr{\"{u}}neis]{Irmler2019a}
Irmler,~A.; Gallo,~A.; Hummel,~F.; Gr{\"{u}}neis,~A. {Duality of Ring and Ladder Diagrams and Its Importance for Many-Electron Perturbation Theories}. \emph{Phys. Rev. Lett.} \textbf{2019}, \emph{123}, 156401\relax
\mciteBstWouldAddEndPuncttrue
\mciteSetBstMidEndSepPunct{\mcitedefaultmidpunct}
{\mcitedefaultendpunct}{\mcitedefaultseppunct}\relax
\EndOfBibitem
\bibitem[Loos \latin{et~al.}(2018)Loos, Romaniello, and Berger]{Loos2018}
Loos,~P.~F.; Romaniello,~P.; Berger,~J.~A. {Green Functions and Self-Consistency: Insights from the Spherium Model}. \emph{J. Chem. Theory Comput.} \textbf{2018}, \emph{14}, 3071--3082\relax
\mciteBstWouldAddEndPuncttrue
\mciteSetBstMidEndSepPunct{\mcitedefaultmidpunct}
{\mcitedefaultendpunct}{\mcitedefaultseppunct}\relax
\EndOfBibitem
\bibitem[{Van Setten} \latin{et~al.}(2015){Van Setten}, Caruso, Sharifzadeh, Ren, Scheffler, Liu, Lischner, Lin, Deslippe, Louie, Yang, Weigend, Neaton, Evers, and Rinke]{VanSetten2015}
{Van Setten},~M.~J.; Caruso,~F.; Sharifzadeh,~S.; Ren,~X.; Scheffler,~M.; Liu,~F.; Lischner,~J.; Lin,~L.; Deslippe,~J.~R.; Louie,~S.~G.; Yang,~C.; Weigend,~F.; Neaton,~J.~B.; Evers,~F.; Rinke,~P. {GW100: Benchmarking G0W0 for Molecular Systems}. \emph{J. Chem. Theory Comput.} \textbf{2015}, \emph{11}, 5665--5687\relax
\mciteBstWouldAddEndPuncttrue
\mciteSetBstMidEndSepPunct{\mcitedefaultmidpunct}
{\mcitedefaultendpunct}{\mcitedefaultseppunct}\relax
\EndOfBibitem
\bibitem[Bruneval \latin{et~al.}(2016)Bruneval, Rangel, Hamed, Shao, Yang, and Neaton]{Bruneval2016a}
Bruneval,~F.; Rangel,~T.; Hamed,~S.~M.; Shao,~M.; Yang,~C.; Neaton,~J.~B. {MOLGW 1: Many-body perturbation theory software for atoms, molecules, and clusters}. \emph{Comput. Phys. Commun.} \textbf{2016}, \emph{208}, 149--161\relax
\mciteBstWouldAddEndPuncttrue
\mciteSetBstMidEndSepPunct{\mcitedefaultmidpunct}
{\mcitedefaultendpunct}{\mcitedefaultseppunct}\relax
\EndOfBibitem
\bibitem[{Te Velde} and Baerends(1991){Te Velde}, and Baerends]{TeVelde1991}
{Te Velde},~G.; Baerends,~E.~J. {Precise density-functional method for periodic structures}. \emph{Phys. Rev. B} \textbf{1991}, \emph{44}, 7888--7903\relax
\mciteBstWouldAddEndPuncttrue
\mciteSetBstMidEndSepPunct{\mcitedefaultmidpunct}
{\mcitedefaultendpunct}{\mcitedefaultseppunct}\relax
\EndOfBibitem
\bibitem[Spadetto \latin{et~al.}(2023)Spadetto, Philipsen, F{\"{o}}rster, and Visscher]{Spadetto2023}
Spadetto,~E.; Philipsen,~P. H.~T.; F{\"{o}}rster,~A.; Visscher,~L. {Toward Pair Atomic Density Fitting for Correlation Energies with Benchmark Accuracy}. \emph{J. Chem. Theory Comput.} \textbf{2023}, \emph{19}, 1499--1516\relax
\mciteBstWouldAddEndPuncttrue
\mciteSetBstMidEndSepPunct{\mcitedefaultmidpunct}
{\mcitedefaultendpunct}{\mcitedefaultseppunct}\relax
\EndOfBibitem
\bibitem[Monzel \latin{et~al.}(2023)Monzel, Holzer, and Klopper]{Monzel2023}
Monzel,~L.; Holzer,~C.; Klopper,~W. {Natural virtual orbitals for the GW method in the random-phase approximation and beyond}. \emph{J. Chem. Phys.} \textbf{2023}, \emph{158}, 144102\relax
\mciteBstWouldAddEndPuncttrue
\mciteSetBstMidEndSepPunct{\mcitedefaultmidpunct}
{\mcitedefaultendpunct}{\mcitedefaultseppunct}\relax
\EndOfBibitem
\bibitem[Gunnarsson \latin{et~al.}(2016)Gunnarsson, Sch{\"{a}}fer, Leblanc, Merino, Sangiovanni, Rohringer, and Toschi]{Gunnarsson2016}
Gunnarsson,~O.; Sch{\"{a}}fer,~T.; Leblanc,~J.~P.; Merino,~J.; Sangiovanni,~G.; Rohringer,~G.; Toschi,~A. {Parquet decomposition calculations of the electronic self-energy}. \emph{Phys. Rev. B} \textbf{2016}, \emph{93}, 245102\relax
\mciteBstWouldAddEndPuncttrue
\mciteSetBstMidEndSepPunct{\mcitedefaultmidpunct}
{\mcitedefaultendpunct}{\mcitedefaultseppunct}\relax
\EndOfBibitem
\end{mcitethebibliography}

\end{document}